\documentclass[a4paper,10pt]{article}
\usepackage{jheppub}
\usepackage[]{array}
\usepackage[]{amsmath}
\usepackage[]{amssymb}
\usepackage[]{mathrsfs}
\usepackage[]{tensor}
\usepackage[symbol]{footmisc}
\newcommand{\del}{\partial}

\newcommand{\tr}{\textrm{tr}}

\newcommand{\be}{\begin{eqnarray}}
\newcommand{\ee}{\end{eqnarray}}
\newcommand{\0}{\nonumber}
\newcommand{\Gam}{\mathbf{\Gamma}}
\newcommand{\R}{\mathbf{R}}

\newcommand{\J}{\mathbf{J}}
\newcommand{\Q}{\mathbf{Q}}
\newcommand{\La}{\Lambda}
\newcommand{\LL}{\mathbf{L}}

\newcommand{\X}{\mathbf{\Xi}}
\newcommand{\thesy}{\boldsymbol{\theta}}
\newcommand{\ome}{\boldsymbol{\omega}}
\newcommand{\Th}{\mathbf{\Theta}}
\newcommand{\CS}{\mathbf{\Upsilon}}
\newcommand{\nc}{^\mathrm{nc}}

\newcommand{\Sig}{\mathbf{\Sigma}}

\newcommand{\al}{\boldsymbol{ \alpha}}

\begin{document}
\begin{flushright}
{SISSA 19/2012/EP \\ ZTF-12-1}\\
{arXiv: 1207.6969 [hep-th]}\\ \today
\end{flushright}
\vskip 1cm
\begin{center}
{\LARGE{\bf  Gravitational Chern-Simons terms and\\[3mm] black hole entropy. Global aspects.}}
\vskip 1cm

{\Large L.~Bonora$^{a}$, M.~Cvitan$^b$, P.~Dominis Prester$^c$, S.~Pallua$^b$, I.~Smoli\'c$^b$}\\
{}~\\
\quad \\
{\em ~$~^{a}$International School for Advanced Studies (SISSA/ISAS),}\\
{\em Via Bonomea 265, 34136 Trieste, and INFN, Sezione di Trieste, Italy}
{}~\\
\quad \\
{\em ~$~^{b}$Physics Department, Faculty of Science,}\\
{\em University of Zagreb, p.p.~331, HR-10002 Zagreb, Croatia}
 {}~\\
\quad \\
{\em ~$~^{c}$ Department of Physics, University of Rijeka,}\\
{\em  Omladinska 14, HR 51000 Rijeka, Croatia}\\
\vskip 1cm
Email: bonora@sissa.it, mcvitan@phy.hr, pprester@phy.uniri.hr, pallua@phy.hr, ismolic@phy.hr

\end{center}

\vskip 2cm {\bf Abstract}. We discuss the topological and global gauge properties of the formula for a black hole entropy due to a purely gravitational Chern-Simons term. We study under what topological and geometrical conditions this formula is well-defined. To this end we have to analyze the global properties of the Chern-Simons term itself and the quantization of its coupling. We show that in some cases the coupling quantization may interfere with the well-definiteness of the entropy formula.

\vskip 1cm 

 \vfill\eject
 
\section{Introduction}

In \cite{Tachi,BCDPS1} a formula was proposed for the modification of the entropy of a black hole
due to the addition of a gravitational Chern-Simons term in the action. The method used is very akin
to the one used by Wald, \cite{Wald1} (see also \cite{Iyer,JKM,Wald2,Wald3,Racz1,Racz2}), the covariant phase space formalism. In order to bypass formal obstacles in the derivation we were obliged in \cite{BCDPS1} to use a particular coordinate system (a Kruskal- type system of coordinates), which led us to a formula for the entropy expressed in such coordinates. Taking this as a starting point it was possible to covariantize it, i.e. to prove that there exists a covariant expression of the entropy that reduces to the local one when the Kruskal-type coordinates are chosen. This however is not enough because it guarantees covariance under local coordinate transformations (and normal bundle infinitesimal gauge transformations), but not necessarily under global transformations. In fact the problem is more general than this. Also the Chern-Simons term\footnote{Throughout the paper when speaking of Chern-Simons (CS) Lagrangian terms we refer to purely gravitational CS terms.} introduced in the action has the same problem: as is was written down in \cite{Tachi,BCDPS1} it is not well defined when the space-time topology is nontrivial. In fact all we did in \cite{BCDPS1} is certainly valid if the space-time topology is trivial, i.e. if the space-time is homeomorphic to flat Minkowski. It should be clear that if, faced to a black hole solution, we are not able to define a fully (local and global) covariant entropy formula, that solution is not physically acceptable. 

In this paper we intend to study the effect of global diffeomorphisms and global gauge transformations both on the CS Lagrangian terms and on the CS entropy formula (which also has the form of the CS term) and determine under what conditions they are well defined. We find out that the global formula for CS terms, as is well known, may differ from the local one. We stress that we want to apply the entropy formula to black holes with $(D-2)$-dimensional event horizons, with topologies not restricted to $(D-2)$-sphere. We shall a priori only assume that the horizon section (more precisely, bifurcation surface on the horizon) is momeomorphic to a compact oriented $(D-2)$-dimensional manifold without boundary (this includes black objects like, e.g., black rings or black saturns, which are known to exist in dimensions higher then four). The term ``black hole'' should be understood with this meaning in the rest of the paper. 

Generally speaking, a global definition of the CS Lagrangian term requires that the corresponding coupling be quantized. As we shall see, this may interfere with the definition of entropy. Proceeding further we analyze under what condition the formula for CS entropy is fully covariant under global gauge transformations and is free of topological ambiguities. This is not always the case. If certain triviality conditions are not satisfied the formula has to be replaced by a global one. We have also figured out situations in which the entropy formula may not be globally well defined. As far 
as the equations of motion are concerned, they do not change, so that derivations in 
\cite{Solodukhin:2005ns,BCDPS1, BCDPS2} remain valid.

The main interest of the paper is on CS in dimensions higher than 3. But, occasionally and for the purpose of comparison, we will analyze also the case $D=3$ (for this case there is already a remarkably large literature, see \cite{DJT1,DJT2,Witten,Witten2,W,DW,Banados,DT,WittenM,SolodukhinCS,Perez,Kraus,Park,Miskovic}).

The paper is organized as follows. In section 2 we update the formalism of \cite{BCDPS1} using the spin connection instead of the affine connection and in section 3 we repeat the derivation of the CS entropy within this new formalism and describe in detail the underlying geometry. Section 4 is devoted to the
globalization of CS terms both in their Minkowski and Euclidean versions. This requires in general a discrete coupling constant. In section 5 we analyze the global properties of the black hole entropy formula and introduce the modification anticipated above when a nontrivial geometry is involved.

\section{Some properties of CS terms and the spin connection}
\label{app:spinconn}

In this section we would like to recall some definitions and well-known properties of CS terms which were not explicitly spelled out in detail in \cite{BCDPS1}. In \cite{BCDPS1} we defined the CS term
in a $D=2n-1$-dimensional space-time $X$, $\CS^{(n)}_{\mathrm{CS}}(\Gam)$, by (formally) going to $D+1$ dimensions via the relation
\be\label{CS}
P_n(\R,\ldots,\R)= d \CS^{(n)}_{\mathrm{CS}}(\Gam)
\ee
where $\R=d\Gam +\Gam\Gam$, and $\Gam$ is the one-form of Christoffel symbols. This implies
\be\label{LCS}
\CS^{(n)}_{\mathrm{CS}}(\Gam) = \lambda n \int_0^1 dt \ P_n (\Gam, \R_t, \dots, \R_t)
\ee
The coupling $\lambda$ will be set  $\lambda=1$, for the time being, and turned on later. 
The adjoint-invariant symmetric polynomial $P_n$ was defined with respect to the Lie algebra $SO(2n-2,1)$.
They are in fact symmetrized traces of $n$ elements of this Lie algebra.
However when writing $P_n (\Gam, \R_t, \dots, \R_t)$ it is more natural to interpret $P_n$ as symmetrized
traces in the Lie algebra of the linear group $GL(2n-1,{\mathbb R})$. The reason is that the one-form matrix $\Gam$ is not antisymmetric. There is of course a difference between the $P_n$'s valued in two different Lie algebras. We notice
that since $\R$ is an antisymmetric two-form matrix, $P_n(\R,\ldots,\R)=0$ for odd $n$, no matter whether we interpret
$P_n$ as symmetrized traces in the Lie algebra of $SO(2n-2,1)$ or $GL(2n-1,{\mathbb R})$. However, for instance,
we have $\tr ({\Gam})=\frac 12  d \log ({\det g})$, $g$ being the Riemannian metric. The general relation is contained in a theorem by Chern and Simons,\cite{CS}. Let us consider the frame bundle $LM$ over our $2n-1$ dimensional space-time $X$ with structure group $GL(2n-1,{\mathbb R})$; let $\thesy$ be a linear connection with curvature $\Th$ and $TP_n(\thesy)= n \int_0^1 dt \ P_n (\thesy, \Th_t, \dots, \Th_t)$ the relevant transgression formula (Chern-Simons term). If $\thesy$ restricts to a connection in a $O(2n-1)$ subbundle (i.e. if the connection is metric), then, for odd $n$,  $P_n(\Th,\ldots,\Th)=0$ and $TP_n(\thesy)$ is exact. For this reason we concentrate on the cases in which $n$ is an \emph{even} integer.

Throughout \cite{BCDPS1} we have in fact interpreted $P_n$ in $\CS^{(n)}_{\mathrm{CS}}(\Gam)$
as relevant to $GL(2n-1,{\mathbb R})$ rather then to $SO(2n-2,1)$. This is irrelevant for our derivations and results in \cite{BCDPS1} because, there, we used only the general Lie-algebraic properties of the $P_n$ polynomials, without reference to a specific Lie algebra. However it is interesting to formulate the problem in terms of an $SO(2n-2,1)$ valued connection and the relevant $P_n$. This means passing from the affine connection $\Gam$ to the spin (Cartan) connection $\hat\al$. The two are related to each other in the well--known way
\be
\Gam =E^{-1}d E+ E^{-1} \hat\al E \label{Gammaalpha}
\ee
where $E=\{E^a{}_{\mu}\}$ is the vielbein matrix. In the following we denote by $\mu,\nu,..=0,\ldots,D-1$ generic world indices and by $a,b,\ldots=0,\ldots,D-1$ generic flat indices.  For curvatures we have 
$\R(\Gam)= E^{-1} \R(\hat\al) E $. We denote simply $\R(\Gam)=\R$ and $\R(\hat\al)= \hat \R$. 

The summation convention is lower left - upper right. For instance
\be
\hat\al^{ab}{}_\mu= E^a{}_\nu \, \partial_\mu E^{b\nu} + E^a{}_\nu \, \Gam_{\mu\sigma}^\nu \, E^{b\sigma}\0
\ee
From this we extract the metricity equation
\be
\hat \nabla_\mu E^{b\lambda} \equiv \partial_\mu  E^{b\lambda} + \Gam_{\mu\nu}^{\lambda} \, E^{b\nu} +\hat\al^b{}_c{}_\mu \, E^{c\lambda} = 0 \label{metric}
\ee

The spin connection will allow us to derive a covariant formula for the entropy in a more geometrical way. 
Our idea is to use 
\be\label{LCSa}
\CS^{(n)}_{\mathrm{CS}}(\hat\al) = n \int_0^1 dt\  P_n (\hat\al, \hat\R_t, \dots, \hat\R_t)
\ee
instead of (\ref{LCS}).
 
In (\ref{LCSa}) $P_n$ represents a symmetric trace over flat indices and antisymmetric generators. In (\ref{LCS}) case $P_n$ represents the symmetric trace over world indices. 
Due to (\ref{Gammaalpha}) the two expressions are not the same. To find an explicit relation between the two 
one must proceed as follows (another approach, valid only locally, is discussed in Appendix A). The transformation (\ref{Gammaalpha}) coincides with a global $\Lambda$
gauge transformation of the connection $\hat\al$ 
\be \label{globalgauge}
\hat \al \longrightarrow \Lambda^{-1} d \Lambda +   \Lambda^{-1}\hat\al\Lambda 
\ee
with the formal replacement $\La \rightarrow E$. In Appendix B one can find transformation formulas for CS terms under global gauge transformations. For instance, one has
\be
\CS^{(2)}_{\mathrm{CS}}(\hat\al)-\CS^{(2)}_{\mathrm{CS}}(\Gam)=
+\frac 13 P_2(E dE^{-1}, E dE^{-1}E dE^{-1})-
 d\,P_2(\Gam,E dE^{-1})\label{CS2a-CS2g}
\ee
and in general
\be
 \CS^{(n)}_{\mathrm{CS}}(\hat\al)-\CS^{(n)}_{\mathrm{CS}}(\Gam)
&=&- \frac{\Gamma(n)^2}{\Gamma(2n)} \, P_n(E dE^{-1},dE dE^{-1}, \ldots, dE dE^{-1})\label{CSna-CSng}\\
&&-d\Bigl[ \frac{\Gamma(n)^2}{\Gamma(2n-1)}\, P_n(\Gam,E dE^{-1},dE  dE^{-1}, \ldots, dE dE^{-1})+\ldots\Bigr]\0
\ee
where dots denote other exact terms.

The two CS terms differ by a topological term (the first term in the RHS is closed but not exact in $(D+1)$-dimensional setting) which does not affect the equations of motion, and by a total derivative of local terms. The problem we would like to clarify next is what are the consequences for our analysis, if any, of replacing $\CS^{(n)}_{\mathrm{CS}}(\Gam)$ with $\CS^{(n)}_{\mathrm{CS}}(\hat\al)$.

\section{The covariant phase space formalism for the spin connection}
\label{app:covphasespace}

We would like now to briefly summarize the derivation of the entropy formula in terms of the spin connection formalism. Repeating the procedure of \cite{BCDPS1} 
\be\label{deltaLCS}
\delta  \CS^{(n)}_{\mathrm{CS}}(\hat\al) = n \, P_n (\delta \hat \al , \hat\R^{n-1}) + d\Th\nc(\hat\al, \delta\hat\al) 
\ee
where
\be\label{Thetanc}
\Th\nc \, \equiv \, - n(n-1) \, \int_0^1 dt\  P_n (\hat\al, \delta \hat\al_t, \hat\R_t^{n-2})
\ee

Let us focus on $\delta \hat \al $. After some work we find
\be
\delta \hat \al_\mu^{ab}&=& \frac 12 \left((\hat\nabla_\sigma \delta E_\mu^a) E^{b\sigma} - (\hat\nabla_\sigma \delta E_\mu^b) E^{a\sigma}- (\hat\nabla_\sigma \delta E^{a\tau} ) E^{b\sigma} g_{\mu\tau} + (\hat\nabla_\sigma \delta E^{b\tau} ) E^{a\sigma} g_{\mu\tau}\right)\label{deltaalpha}\\
&=& \frac 12 \left(E^{a\tau}E^{b\sigma} - E^{b\tau}E^{a\sigma}\right) \hat \nabla_\sigma \delta g_{\mu\tau}\0\\
 &=& \frac 12 E^{a\tau}E^{b\sigma} \left(\hat \nabla_\sigma \delta g_{\mu\tau} -\hat \nabla_\tau \delta g_{\mu\sigma}\right) \0
\ee
Let us consider now the term (which contributes to the equation of motion)
\be
 P_n(\delta\hat\al,\hat\R^{n-1})&=& P_n\left( \frac 12 E_a^\tau E^{b\sigma} \left( \hat \nabla_\sigma \delta g_{\mu\tau} -\hat \nabla_\tau \delta g_{\mu\sigma}\right), E_\rho^a E_b^\lambda (\R^{n-1})^\rho{}_\lambda\right)\label{Pnhatalpha}\\
&=& P_n\left( \frac 12 g^{\sigma\tau}\left( \hat \nabla_\sigma \delta g_{\mu\lambda} -\hat \nabla_\lambda \delta g_{\mu\sigma}\right), (\R^{n-1})_\tau{}^\lambda\right)\0
\ee
This is the same form taken by the analogous term in section 2 of \cite{BCDPS1}. Assuming that the rest of the action
depends only on the metric (and not explicitly on the vielbein) and since
\be
\frac {\delta}{\delta E^a_\mu} = 2 {E_{a\nu}} \frac {\delta}{\delta g_{\mu\nu}}\label{deltaE}
\ee
one can see that the equation of motion obtained by the variation with respect to $E$ is the same as the equation of motion obtained by the variation with respect to the metric.  

Next let us come to the symmetry operations. In the metric formalism we write $\delta_\xi$ to represent the overall symmetry (the diffeomorphisms) we are interested in. In the present vielbein formalism the transformation
$\delta_\xi$ does not encompass all possible symmetry operations, we have to include also the local Lorentz
transformations, represented by a local antisymmetric matrix $L^{ab}$. Therefore in the following
we will write
\be
\delta_{\xi,L} = \delta_{\xi} + \delta_L\label{deltaxiL}
\ee
So
\be
\delta_{\xi,L} \hat\al = \delta_{\xi} \hat\al+ \delta_L \hat\al= L_\xi  \hat\al+ D L\label{deltaxiLalpha}
\ee
where $L_{\xi}=i_{\xi}d+di_{\xi}$ and $DL= dL +[\hat\al ,L]$. 
In particular the operation $\hat\delta_\xi$ will be replaced by $\hat \delta_L$, with
\be
\hat \delta_L \hat \al = dL\label{hatdeltaL}
\ee

The request of covariance is for any $\xi$ and any $L$. For the reasons which will be apparent soon, we now fix $L$ to be 
\be
L^{ab} = \frac 12 \left( E^{a\mu} \partial_\mu \xi^\nu E^b_\nu- E^{b\mu} \partial_\mu \xi^\nu E^a_\nu\right)
\label{Lab}
\ee
In effect, (\ref{Lab}) will allow us to use the logic of derivation elaborated in detail in \cite{BCDPS1}, in a straightforward way. For this reason, we shall just present the main points of it.

We start from CS contribution to a current corresponding to the variation (\ref{deltaxiL}) and (\ref{Lab}).
\be \label{JxiL}
\J_{\xi,L} = \Th^{\textrm{cov}}_{\xi}+\Th^{\textrm{nc}}_{\xi,L} - i_{\xi}\LL_{\mathrm{pCS}}-\X_{L} 
\ee
where by $\Th\nc_{\xi,L}$ we mean that in (\ref{Thetanc}) $\delta\hat\al$ has been replaced
by (\ref{deltaxiLalpha}). Also,
\be\label{Xialpha}
\X_L (\hat\al) = n(n-1) \int_0^1 dt \, (t-1) P_n (dL, \hat \al, \hat\R_t^{n-2})
\ee
The current $\J_{\xi,L}$ is conserved on-shell, $d\J_{\xi,L} \approx 0$. The symbol ``$\approx$'' emphasizes that equations of motion where assumed. In the rest of the derivation in this subsection equations of motion are assumed in all expressions, so ``$=$'' should be understood as $\approx$.  

Using the methods of \cite{BCDPS1} we obtain
\begin{equation}
\J_{\xi,L} = d \Q_{\xi,L}
\end{equation}
where the charge $\Q_{\xi,L}$ is of the form
\begin{equation}
\Q_{\xi,L} = \Q^{(1)}_{\xi,L} + \Q^{(0)}_{\xi,L}
\end{equation}
where
\be \label{Q1}
\Q^{(1)}_{\xi,L}(\hat\al) = n(n-1)\int_0^1dt P_n(L,\hat\al,\hat\R_t^{n-2})
\ee
and $\Q^{(0)}_{\xi,L}$ will not be important because it vanishes when $\xi = 0$, which happens for Killing horizon generator on the bifurcation surface.

The Hamiltonian corresponding to the variation (\ref{deltaxiL}) and (\ref{Lab}) is defined by
\begin{equation} \label{Hamom}
\delta H[\xi,L] = \int_\mathcal{C} \boldsymbol{\omega}(\phi, \delta \phi, \delta_{\xi,L} \phi)
\end{equation}
where the $(D-1)$-form $\boldsymbol{\omega}$ is the symplectic current, $\mathcal{C}$ is some Cauchy
surface, $\phi$ denotes collectively all degrees of freedom (``dynamical fields'') in the theory, and 
$\delta$ denotes general variation of fields. It can be shown that the Lagrangian CS term contribution is
\begin{equation}
\delta H[\xi,L] = \int_{\partial\mathcal{C}} \left( \delta \Q_{\xi,L} - \imath_\xi \Th - \Sig_{\xi,L} \right)
\end{equation}
where the CS contribution to $\Sig_{\xi,L}$ is
\be \label{sigmaL}
\Sig_{\xi,L} = -n(n-1)(n-2) \int_0^1 dt \, t(t-1) P_n (dL, \hat\al, \delta \hat\al,\hat \R_t^{n-3})
\ee

We now assume that the solution to the equations of motion is a black hole geometry with a Killing horizon, generated by a vector field we take to be $\xi$ which has a bifurcation surface $\mathcal{B}$, and that the surface gravity $\kappa$ is constant on the horizon. As mentioned before, we assume that  $\mathcal{B}$ is a $(D-2)$-dimensional compact oriented manifold without boundary (not necessarily with the topology of $S^{D-2}$ sphere). An example is given by stationary rotating black holes. We also assume that in this solution all dynamical fields are symmetric, i.e., $\delta_{\xi,L} \phi = 0$, which, by (\ref{Hamom}), enforces $\delta H[\xi,L] = 0$. Using this in (\ref{Hamom}) we obtain
\begin{equation} \label{delHBinf}
\int_{\mathcal{B}} \left( \delta \Q_{\xi,L} - \imath_\xi \Th - \Sig_{\xi,L} \right)
 = \int_{\infty} \left( \delta \Q_{\xi,L} - \imath_\xi \Th - \Sig_{\xi,L} \right)
\end{equation}
This relation has the form of the first law of thermodynamics
\begin{equation}
T \, \delta S = \delta U + \ldots 
\end{equation}
We want to find the expression for the CS contribution to the black hole entropy formula.  This means analyzing the left hand side of (\ref{delHBinf}). In particular we want to integrate the variation of the entropy. First we use the fact that on bifurcating horizon
\begin{equation}
\xi \big|_\mathcal{B} = 0 \quad \Longrightarrow \quad \imath_\xi \Th \big|_\mathcal{B} = 0
\end{equation}
To handle the $\Sig$-term in the left hand side of (\ref{delHBinf}) we use the same trick as in \cite{BCDPS1} - we make our calculations in a particular Kruskal-type coordinate system and at the end covariantize the result. In Appendix \ref{app:Kruskal} we showed (see Eq. (\ref{iBSig0})) that in Kruskal-type coordinates this term also vanishes. Using the familiar expression for the black hole temperature 
\begin{equation}
T = \frac{\kappa}{2\pi}  
\end{equation}
we obtain that the CS contribution to the entropy, evaluated in Kruskal-type coordinates, is
\be \label{SCSpt0}
S_{\mathrm{CS}} = \frac{2\pi}{\kappa} \int_\mathcal{B} \Q^{(1)}_{\xi,L}
 = \frac{2\pi}{\kappa} n(n-1) \int_0^1 dt \int_{\mathcal{B}} P_n(L,\hat\al,\hat\R_t^{n-2})
\ee
Finally, from (\ref{Lab}) and the relation satisfied by Killing generator
\begin{equation}
\partial_\mu \xi^\nu \big|_\mathcal{B} = \kappa \, \epsilon^\nu{}_\mu \;\;,
\end{equation}
where $\epsilon_{\mu\nu}$ is binormal 2-form of $\mathcal{B}$, we obtain
\be \label{SCSpta}
S_{\mathrm{CS}} = 2\pi\, \lambda\, n(n-1) \int_0^1 dt \int_{\mathcal{B}} P_n(\hat\epsilon,\hat\al,\hat\R_t^{n-2})
\ee
where $\hat\epsilon \equiv E \epsilon E^{-1}$ and we have reinserted the coupling $\lambda$. We will focus henceforth on this formula, but before proceeding we need to clarify its geometrical meaning. Most of next section is taken from Appendix D of  \cite{BCDPS1}. We reproduce it here for completeness.

\subsection{Reduction geometry and the entropy formula}

The entropy formula (\ref{SCSpta}) has an interesting geometrical interpretation. 
In order to appreciate it it is useful to review the geometrical setting underlying the problem we are studying, see \cite{KN}, vol.II. The geometry is that of an asymptotically Minkowski space-time manifold  $X$ with a codimension 2 submanifold ${\mathcal{B}}$. We have $O(X)$, the bundle of orthonormal frames on $X$ with structure group
$SO(D-1,1)$ and $O({\mathcal{B}})$ the bundle of orthonormal frames on ${\mathcal{B}}$ with structure group
$SO(D-2)$. We consider also the bundle of adapted frames. An adapted frame is a complete set of orthonormal vectors which are either tangent or orthogonal to ${\mathcal{B}}$. They form a principal bundle $O(X,{\mathcal{B}})$ with structure group
$SO(1,1)\times SO(D-2)$. To complete the description we have the bundle
of normal frames $ON({\mathcal{B}})$ with structure group
$SO(1,1)$ and the embedding $i$: $O(X,{\mathcal{B}}) \stackrel{i}{\longrightarrow} O(X)$.
For convenience, let us denote by ${\mathfrak h}$ and ${\mathfrak k}$ the Lie algebras
of $SO(D-2)$ and $SO(1,1)$, respectively.

Concerning the connections, let us repeat that $\Gam$ is a connection of the linear frame bundle $LX$. Every
metric connection in $LX$ is in one-to-one correspondence with a connection in $O(X)$ (see \cite{KN}, vol.I, ch. 4, $\S$ 2). The connection $\hat\al$ is a connection in $O(X)$.

In (\ref{SCSpta}) it is understood that the forms in the integrand are pulled back from $X$
to ${\mathcal{B}}$. Now by pulling back a generic connection $\hat\al$ of $O(X)$ through $i$, we do not get a connection, unless we restrict to the components in ${\mathfrak h}$+
${\mathfrak k}$. If so, the connection splits into
$\hat\al_t+\hat\al_\perp$, that is a connection  $\hat\al_t$ in $O({\mathcal{B}})$ with values
in $\mathfrak h$ and a connection $\hat\al_\perp$ in $ON({\mathcal{B}})$ with values in $\mathfrak k$ (see \cite{KN}, vol.II, ch. VII). The geometry of the problem is defined by the presence of the surface ${\mathcal{B}}$ with its tangent and normal directions, thus the just considered reduction of a connection pulled back from $X$,
is natural in this scheme. But once we replace in (\ref{SCSpta}) the connection $\hat\al_t+\hat\al_\perp$, with values in the direct sum ${\mathfrak h}$+ ${\mathfrak k}$, the presence
of the binormal $\epsilon$ maps out the ${\mathfrak h}$ components and only the components
along ${\mathfrak k}$ (the Lie algebra of the normal frame bundle with structure group $SO(1,1)$) survive.

There is also another reason why this simplification occurs. Once we reduce to the direct sum ${\mathfrak h}$+ ${\mathfrak k}$, the polynomial $P_n$ splits into the sum of the polynomials $P_n^{({\mathfrak h})}$ over ${\mathfrak h}$ and a polynomial  $P_n^{({\mathfrak k})}$ over ${\mathfrak k}$. The first polynomial vanishes
because it is a trace over the Lie algebra of $SO(D-2)$ and $D=2n-1$, with $n$ even. At this point we
are left with an Abelian connection and we can easily integrate over $t$. If we call $\ome$ one of the two identical components of the form matrix $\hat\al_t$, we get easily
the formula obtained in \cite{BCDPS1}
\be \label{SCSp}
S_{\mathrm{CS}} = 4\pi\,\lambda \, n  \int_{\mathcal{B}}  \ome \,(d\ome)^{n-2}
\ee

To view the situation in more detail let us introduce the following conventions (in this regard see also 
Appendix \ref{app:eugCSent}. Following  \cite{Carter:2000wv}, we will denote
by $A,B,..=2, \ldots, D-1$ flat tangent indices in ${\mathcal{B}}$ and by $X,Y,\ldots=0,1$ normal flat indices (these two sets of indices are collectively denoted by $a,b$), and introduce
adapted vielbein $\imath_A{}^\mu$ and $\lambda_X{}^\mu$ (they are particular cases of $E_a^\mu$) , such that
\be
q^{\mu}{}_\nu = \imath_A{}^\mu \imath^A{}_\nu,\quad\quad h^\mu{}_\nu = \lambda_X{}^\mu \lambda^X{}_\nu\label{qandh}
\ee
One can show in particular that, since $ \lambda_X{}^\mu\lambda_{Y\mu} =\eta_{XY}$ ($\eta$ denotes the flat Minkowski metric), one can make the following identifications 
\be
\lambda_0{}^\mu= \frac {n^\mu-l^\mu}{\sqrt{2}},\quad\quad \lambda_1{}^\mu= \frac {n^\mu+l^\mu}{\sqrt{2}}\label{lambdanl}
\ee 
with reference to the null vectors introduced in the previous Appendix.
Then, it is easy to show that
\be
\epsilon_{\mu\nu} E_a{}^\mu E_b{}^\nu=\eta_{1a} \eta_{0b}-\eta_{1b} \eta_{0a} \label{eEE}
\ee 
where $\eta$ is the flat Minkowski metric. Thus, for instance,
\be
\tr (\hat\epsilon \hat\al_\perp)=\epsilon_{\mu\nu} E_a{}^\mu E_b{}^\nu (\hat\al_\perp)^{ab}
 = 2 \hat\al_\perp^{01}\equiv 2\,\omega\label{alpha01}
\ee
and likewise
\be \label{R01}
\tr (\hat\epsilon \hat\R_\perp^{n-2} )= 2 (\hat \R_\perp^{01})^{n-2}
\ee
for the curvature.  Therefore in this approach we obtain the same formulas as in \cite{BCDPS1}    with
$\Gam$ and $\R$ replaced by $\hat\al_\perp$ and its curvature $\hat \R_\perp$. It is understood that all the forms 
are pulled back to ${\mathcal{B}}$, which can be achieved on components by contracting the form index with the $q$ projector: for instance the intrinsic component of the pulled back $\hat\al_\perp$ is $q_\mu{}^\nu (\hat\al_\perp)_\nu$.

It is now convenient to compare the normal bundle connection  with the one introduced in \cite{Carter:2000wv},
\be
\varpi_\mu{}^\nu{}_\rho = h_\sigma{}^\nu \lambda^X{}_\rho \bar \nabla_\mu \lambda_X{}^\sigma, \quad {\rm where}\quad 
\bar \nabla_\mu= q^\nu{}_\mu \nabla_\nu\label{varomega}
\ee
Using $\nabla E_a^\mu= -\hat\al_a{}^b E_b^\mu$ we can rewrite
\be
\varpi_\mu{}^\nu{}_\rho= q_\mu{}^\sigma \epsilon_\rho{}^\nu (\hat\al_n)_\sigma^{01}\label{varpi2}
\ee
Saturating with $\epsilon_\nu{}^\rho$ we obtain precisely the RHS of (\ref{alpha01}).

On the other hand, inserting (\ref{lambdanl}) into (\ref{varomega}) one finds
\be 
\varpi_\mu{}^\nu{}_\rho = -\epsilon^\nu{}_\rho n_\tau \bar \nabla_\mu \ell^\tau \label{varpi3}
\ee
Saturating with $\epsilon^\rho{}_\nu$ and dividing by 2, we get precisely the definition (D.10) in \cite{BCDPS1}.

Finally a comment about gauge transformations in the normal frame bundle. They are valued in $SO(1,1)$ and act on $l_0,l_1$ as follows
\be
\left(\begin{matrix} l_0 \\ l_1\end{matrix} \right) \rightarrow \left(\begin{matrix} \cosh f & \sinh f \\ \sinh f & \cosh f \end{matrix} \right)\left(\begin{matrix} l_0 \\ l_1\end{matrix} \right)\label{so11}
\ee
where $f$ is a local function. Using again (\ref{lambdanl}), it is easy to see that they act on $n,l$ as a rescaling
\be
n\rightarrow e^f n,\quad\quad l\rightarrow e^{-f} l\label{transfnl}
\ee
Under this rescaling, $\ome$ transforms as
\be \label{rescaling}
\ome\rightarrow \ome - d f
\ee

\vskip 1cm

{\bf Remark 1}. From eq.(\ref{so11}) we see that the generator corresponding to an infinitesimal gauge transformation is represented 
by the matrix $L=\left(\begin{matrix} 0&1 \\ 1&0\end{matrix} \right)$, which is symmetric. 
The correspondence with the matrix entries is $L_{00}=L_{11}=0$ and $L_{01}=L_{10}=1$. However, in order to saturate the indices inside the trace in $P_n$ we have to raise one of the two with the flat Minkowski metric, so $L_0{}^1=-L_1{}^0$.

{\bf Remark 2}. Eq.(\ref{SCSp}) is strikingly similar to the volume form of a contact manifold, with 
$\ome$ playing the role of contact form. If this was the case $S_{\mathrm{CS}}$ would be proportional to the volume of $\cal B$. We notice however that an essential condition for $\ome$ to be identified with
a contact form is that it be nowhere vanishing, a condition that our context does not in general allow to grant. We shall say more on this and some other related issues in the forthcoming paper \cite{BCDPSp}.

\vskip 1cm

Let us end this section with a final remark concerning the consequences of a Wick rotation on eq.(\ref{SCSp}).
From (\ref{transfnl},\ref{rescaling}) we see that the gauge transformation on $n$ and $l$ takes values in ${\mathbb R}$,
the group of real numbers. If we make a Wick rotation, the structure group of the normal bundle becomes
$SO(2)$ and the entropy formula becomes
\be \label{SCSpb}
S_{\mathrm{CS}} = -2\pi\,i\,\lambda\,  n \int_{\mathcal{B}} \tr \left(\hat \epsilon\, \hat\al_\perp(\hat\R_\perp)^{n-2} \right)
\ee
where $\hat\al_\perp$ takes values in the Lie algebra of $SO(2)$ (for the appearance of the imaginary unit
see below). If we set $(\hat\al_\perp)^{10}=-(\hat\al_\perp)^{01}\equiv \al$, we can rewrite (\ref{SCSpb})
as
\be \label{SCSpeu}
S_{\mathrm{CS}} = 4\pi\,i\,\lambda\,  n \int_{\mathcal{B}} \al (d\al)^{n-2} 
\ee
Since a generic local $SO(2)$ gauge transformation is represented by the matrix
\be \label{gaugetrso2}
\Lambda =  \left( \begin{matrix} \cos g & \sin g \\ -\sin g & \cos g \end{matrix} \right)
\ee
where $g$ is a real-valued function on ${\mathcal{B}}$, 
the gauge transformation $\hat\al_\perp\to \Lambda^{-1} (d+\hat\al_\perp)\Lambda$ implies 
for the real-valued form $\al$
\be \label{gaugeso2}
\al \to \al + d g 
\ee
As one can see, the gauge transformations of $\al$ and $\ome$ take the same form.

As a final comment here, let us note that both (\ref{SCSp}) and (\ref{rescaling}) (Minkowski case), and 
(\ref{SCSpeu}) and (\ref{gaugeso2}) (Euclidean case) suggest that CS entropy term is itself an (Abelian) CS term in vector bundle associated to principal bundle with base space $\mathbf{B}$, gauge group $SO(1,1)$ or $U(1)$, and where the fiber of the vector bundle is \emph{one-dimensional}.

\section{Global aspects of the CS term}
\label{app;global}

So far, both in this paper and in \cite{BCDPS1}, we have derived all our results using an essentially local formalism. So, in particular, our entropy formula for CS terms is covariant as long as local coordinate
transformations and local gauge transformations in the normal bundle (\ref{rescaling}) are considered, but we have to ask ourselves whether it is covariant also under global
transformations and more generally: are formulas (\ref{SCSp}) or (\ref{SCSpb}) well-defined in a nontrivial
topological framework? Do we have to change them in the presence of nontrivial topology?
Actually the same question should be asked for the initial gravitational CS term itself
(\ref{LCS}). The question is: how does (\ref{LCS}) change when we consider it in a nontrivial topological setting? Does such a change modify significantly our earlier derivations? 

When global aspects are involved in formulas with a geometrical character a question immediately arises:
should we consider them also in the Euclidean version? Since the Euclidean CS term has the flavor of a topological quantity, it is sensible to ask this question. Considering that Euclidean approach generally offers some powerful methods of analyses, it would be important to include it. Therefore we will consider both Minkowski and Euclidean versions in turn.

We will start from the global aspects of the Lagrangian CS term.

\subsection{Is the CS coupling quantized?}

Let us denote by $I_0$ the gravity action and by $I_{CS}$ the CS action.
The Minkowski path integral of the theory we are interested in is, in general, 
\be
Z[\phi]= \int {\cal D}\phi e^{i (I_0+ I_{CS})}\label{Minkowski}
\ee
where $I_0=\int L_0$ and $I_{CS}= \lambda \int L_{CS}$, where we have extracted explicitly the CS coupling $\lambda$.
 
The Euclidean path integral is 
\be
Z_E[\phi]= \int {\cal D}\phi e^{- (I^{(E)}_0-i I^{(E)}_{CS})}\label{Euclid}
\ee
the label $^{(E)}$ denotes the Euclidean version.
The reason for this is well-known:  $I_{CS}$ contains the totally antisymmetric $\epsilon$ tensor appropriate for the given space-time $X$; thus the Wick rotation $x^0\to i x_E$ operates only twice in  $I_{CS}$, in the measure and in the unique index 0 
appearing in the integrand, so $I_{CS}\to I_{CS}$. Therefore, in the exponent of the path integral, the CS action appears always with an $i$ in front. The action now is $I_0-i I_{CS}= \int(L^{E}_0-i\lambda L^{(E)}_{CS})$. The result  is that the topological $L^{(E)}_{CS}$ has an imaginary coupling. 
If we want to go back to Minkowski we have to replace the results
with their Minkowski form and change back $-i\lambda \to \lambda$. In particular we remark that
in the entropy formula the coupling appears linearly. Therefore the CS entropy in the Euclidean becomes imaginary, going back to Minkowski it returns to its real form.

\subsubsection{The Euclidean version of the CS action}
\label{sssec:euCSact}

Let us consider, in general, a connection 1-form $\mathbf{A}$ in a principal fiber bundle $P(X,G)$ with structure group $G$.
The expression
\be
I^{(E)}_{CS} = \lambda \, n \int_{X} d^D x\,\int_0^1 dt\,
 P_n(\mathbf{A},\mathbf{F}_t,\ldots,\mathbf{F}_t)\label{ICSX} 
\ee
where $D=2n-1$, is not well defined  in general. The reason is the following one. A connection is a one-form
defined on the total space. Thus (\ref{ICSX}) is well defined in the total space $P$, but not on the base $X$. 
When in field theory we write down a formula like this we usually mean that $A$ is pulled back via a local section $\sigma$: $\sigma^* A$ is a local form on the base manifold, but defined in a local patch. Thus the integrand in $I^{(E)}_{CS}$ is defined only in such a local patch. Equivalently we can say that, in general, $A$ has Dirac string singularities. To define $I^{(E)}_{CS}$ globally we proceed in a well-known  way, \cite{DW,W,Witten}.

We take a bounding manifold ${\cal V}$, i.e. a manifold such that $\del{\cal V}=X$, and, using Stokes theorem, we set
\be 
I^{(E)}_{CS} = \lambda \int_{\cal V} d^{2n} x\, P_n(\mathbf{F}) \label{ICSV}
\ee
where $P_n(\mathbf{F}) \equiv P_n(\mathbf{F},\ldots,\mathbf{F})$. This is now a well defined integral in 
${\cal V}$, because $\mathbf{F}$, the curvature, is always a basic form in $X$ if it can be extended by continuity to ${\cal V}$. However the action now depends on the choice of ${\cal V}$, which is clearly 
non-physical. If we integrate on another bounding manifold  ${\cal V}'$ the difference between the two integrals, i.e. the ambiguity of $I^{(E)}_{CS}$, is given by
\be
 \int_{\cal Z} d^{2n} x\, P_n(\mathbf{F}) \label{PkF}
\ee
where ${\cal Z}= {\cal V}-{\cal V}'$ ($-{\cal V}'$ means that ${\cal V}'$ is glued to
${\cal V}$ with the opposite orientation). ${\cal Z}$ is a closed oriented manifold of dimension $2n$. In order for the ambiguity to be harmless it must be that
\be
\lambda \int_{\cal Z} d^{2n} x\, P_n(\mathbf{F})\in 2\pi {\mathbb Z} \label{PkFcond}
\ee
because in this case the exponential in the path integral (\ref{Euclid}) is unchanged. 

It is known that every $P_n(\mathbf{F})$ may be written as a linear combination of terms which are products of traces, so, for practical purposes, we may assume that $P_n$ is given by
\begin{equation} \label{Pntraces}
P_n(\mathbf{F}) = \left( \tr(\mathbf{F}^{k_1}) \right)^{l_1} \cdots \left( \tr(\mathbf{F}^{k_r}) \right)^{l_r}
 \;\;, \qquad \sum_{j=1}^r k_r l_r = n
\end{equation}
where the traces are supposed to be evaluated in the fundamental representation of the relevant group.
Then $\lambda$ is a coupling constant corresponding to the term (\ref{Pntraces}). Every term of the type
(\ref{Pntraces}), after proper normalization, can be written as polynomial made of Pontryagin classes with \emph{integer} coefficients (see Appendix \ref{app:indexthrm}), which means that (\ref{PkF}) must be an integer times normalization constant. It then follows that $\lambda$ must be quantized
\be
\lambda \sim k \in{\mathbb Z}\label{lambdaquant}
\ee
to secure that (\ref{PkFcond}) will be obeyed.

Let us now apply the previous general remarks to our case of interest with base manifold $\mathcal{Z}$ 
of dimension  $2n$, where the relevant bundle
is tangent bundle $T\mathcal{Z}$ associated to principal orthogonal bundle with structure group $SO(2n)$,
so that $F$ will be replaced by $\R$. Details of calculations can be found in Appendix
\ref{app:indexthrm}. Let us first specialize (\ref{Pntraces}) to the case of an ``irreducible'' trace, i.e.,
\begin{equation} \label{Pnirrtr}
P_n(\mathbf{\R}) = \tr(\mathbf{\R}^{n})
\end{equation}
where $\tr$ is the trace in the fundamental representation.
By expressing (\ref{Pnirrtr}) in terms of Pontryagin classes one finds
\begin{equation} \label{irtrintPo}
\frac{1}{2(2\pi)^n} \int_\mathcal{Z} \tr(\mathbf{\R}^{n}) \, \in \, {\mathbb Z}
\end{equation}
which when used in (\ref{PkFcond}) leads to the quantization condition
\begin{equation} \label{lquantPo}
\lambda \, \in \, \frac{1}{2(2\pi)^{n-1}} \, {\mathbb Z} \;. 
\end{equation}

Let us stress the meaning of the quantization condition (\ref{lquantPo}): when $\lambda$ satisfies (\ref{lquantPo}) we are sure that (\ref{PkFcond}) will be satisfied and, thus, the action  
(\ref{ICSV}) will be well defined. 

It should be added that the quantization rule (\ref{PkFcond}) is generic.  
As we shall show in the next subsection, it can be, at least partially, relaxed. But, before, let us remark that our argument concerning the well-definiteness of (\ref{ICSV})
is still largely incomplete. The reason is that it is not guaranteed
that a bounding manifold ${\cal V}$ exists for $X$. There may be obstructions. The mathematical theory that takes care of this kind of problems is {\it cobordism theory}, see \cite{MS,Stong}.
Two manifolds $X_1$ and $X_2$, belonging to a given class, (for instance, that of smooth manifolds) are said to be {\it cobordant} if there exists a smooth manifold ${\cal Y}$ of the same class such that $\del {\cal Y}= X_1-X_2$ (the - in front of $X_2$ means that its orientation is reversed). The relation of being cobordant is symmetric and transitive: it splits manifolds into classes, which can be summed in a
natural way (union of manifolds). Thus they form  abelian groups, which are called
cobordism groups and are denoted by $\Omega_D$ for manifolds of dimension $D$. The zero element of the group denotes the class of manifolds that are cobordant to the empty set, i.e. the class of manifolds that have bounding manifolds. It is clear that, if our manifold $X$ belongs to this class, we are allowed to pass from (\ref{ICSX}) to (\ref{ICSV}).

This is not yet enough, because in passing from (\ref{ICSX}) to (\ref{ICSV}) we have to extend the bundle, where $A$ is defined, from $X$ to ${\cal V}$. In the case we are interested 
in, $A$ is $\hat \al$. The bundle in question is the orthonormal bundle with gauge group $SO(2n-1)$ (as long as we stick to oriented manifolds). The cobordism groups relevant in this case are $\Omega_D^{SO}$.
We have in particular
\be
&&\Omega_1^{SO}=\Omega_2^{SO}=\Omega_3^{SO}=0,\quad \Omega_4^{SO}={\mathbb Z},\quad
\Omega_5^{SO}={\mathbb Z}_2,\quad \Omega_6^{SO}=\Omega_7^{SO}=0,\label{cobSO}\\
&&\Omega_8^{SO}={\mathbb Z}\oplus{\mathbb Z},\quad  \Omega_9^{SO}={\mathbb Z}_2\oplus {\mathbb Z}_2,\ldots\0
\ee
This means that all 3- and 7-dimensional manifolds have bounding manifolds. Other odd dimensional manifolds may not have, depending on what class the manifold belongs to. On the other hand if black holes are studied in a geometry which is asymptotically Minkowski, and the Euclidean version of the asymptotic geometry is a sphere (which represents the one point compactification  of ${\mathbb R}^n$), since any  sphere has bounding manifolds, we can limit ourselves to considering manifolds 
in the zero class of $\Omega_*^{SO}$. In all these cases, that is for a very large class of manifolds, (\ref{ICSV}) is a good representation of (\ref{ICSX}), or, better, for our case,
\be
 \int_{\cal V} d^{2n} x\, P_n(\mathbf{R}) \label{PkR}
\ee
is a good representation of (\ref{LCS}) or (\ref{LCSa}).

\subsubsection{Relaxing the coupling quantization}

As we have anticipated above, the quantization  (\ref{lambdaquant}) is generic. The exact proportionality coefficient in (\ref{lambdaquant}) may depend on the characteristics of the manifold where the theory is defined and on the configuration space of solutions it requires. This is a well-known fact established 
and elaborated in some detail in the special case $n=2$ (i.e., $D=3$), see, e.g., \cite{Witten,W,DW}. In what follows we would like to give some examples for other values of $n$.  
  
Specifically, in this paper, we deal not with some arbitrary gauge theories with $SO(2n)$ gauge group, but with (Euclidean) theories of gravity. In classical gravity the connection is the Levi-Civita connection $\Gam$ (or the spin connection $\hat {\mathbf \alpha}$) and the curvature is the Riemann curvature 2-form $\mathbf{R}$, which can be obtained from nonsingular (i.e., invertible) metric tensor (or vielbein). To emphasize this, in the previous formulas we will make the replacements 
$\mathbf{A} \to \Gam$ and $\mathbf{F} \to \mathbf{R}$. Let us now assume that in the path integral (\ref{Euclid})  only such configurations that are classically well-defined and regular have to be taken into account. This effectively restricts the configuration space, as not all connections allowed in gauge theory are non-singular in classical gravity terms. We can now use the Hirzebruch signature theorem and gain extra information not present in (\ref{irtrintPo}). This time, though, one has to work out the result for each $n$ case by case. Interestingly, for $n = 2,4,6$, using Hirzebruch theorem, the result can be written in compact form:
\begin{equation} \label{irtrintHi}
\frac{1}{2(2\pi)^n} \int_\mathcal{Z} \tr(\mathbf{R}^{n}) \, \in \, (n+1) \, {\mathbb Z}
 \;\;, \qquad n = 2,4,6
\end{equation}
We were not able to extend (\ref{irtrintHi}) to $n > 6$. Using this together with (\ref{PkFcond})
gives us the new quantization condition
\begin{equation} \label{lquantHi}
\lambda \, \in \, \frac{1}{2(2\pi)^{n-1}} \, \frac{\mathbb Z}{n+1}  \;\;, \qquad n = 2,4,6
\end{equation}
which is less restrictive than (\ref{lquantPo}). This is a consequence of the
reduction of the configuration space.

A configuration space may be further reduced by requiring existence of additional structures. For example, we may require the theory to couple to Dirac fermions, thus the base manifold to be a spin manifold. In this case we can use Atiyah-Singer index theorem for the Dirac operator to gain more information. Again, this has to be worked out case by case for each $n$. For $n=2$ one gets
\begin{equation} \label{irtrintASn2}
\frac{1}{8\pi^2} \int_\mathcal{Z} \tr(\mathbf{R}^2) \, \in \, 48 \, {\mathbb Z}
\end{equation}
which, when used in (\ref{PkFcond}), yields the quantization condition
\begin{equation} \label{lquantASn2}
\lambda \, \in \, \frac{1}{4\pi} \, \frac{\mathbb Z}{48} \;. 
\end{equation}
For $n=4$ one gets
\begin{equation} \label{irtrintASn4}
\frac{1}{2(2\pi)^4} \int_\mathcal{Z} \tr(\mathbf{R}^4) \, \in \, 10 \, {\mathbb Z}
\end{equation}
which when used in (\ref{PkFcond}) gives
\begin{equation} \label{lquantASn4}
\lambda \, \in \, \frac{1}{2(2\pi)^3} \, \frac{\mathbb Z}{10} \;. 
\end{equation}
We see that conditions  (\ref{lquantASn2}) and (\ref{lquantASn4}) are less restrictive then (\ref{lquantHi}), a consequence of the additional reduction of the configuration space due to the requirement of the manifold being spin.

A similar analysis may be performed with different choices in (\ref{Pntraces}). In general, the 
corresponding coupling constants will have different quantization conditions. As an illustration, let us instead of (\ref{Pnirrtr}) now take one of reducible products of traces, for example
\begin{equation} \label{Pnredtr}
P_n(\mathbf{R}) = \left( \tr(\mathbf{R}^2) \right)^{n/2} \;\;, \qquad n \ge 4
\end{equation}
To avoid possible confusion, let us denote by $\lambda'$ the coupling constant corresponding to term (\ref{Pnredtr}). We can use Pontryagin classes to obtain
\begin{equation} \label{redtrintPo}
\frac{1}{2^{n/2}(2\pi)^n} \int_\mathcal{Z} \left( \tr(\mathbf{R}^2) \right)^{n/2}
 = \int_\mathcal{Z} \left( p_1(\mathbf{R}) \right)^{n/2}
 = P_1^{n/2}(\mathbf{R}) \, \in \, {\mathbb Z}
\end{equation}
where $p_1$ is the first Pontryagin class and $P_1^{n/2}$ one of Pontryagin numbers (which are always integers). By using (\ref{redtrintPo}) in (\ref{PkFcond}) we obtain the quantization condition
\begin{equation} \label{redlquantPo}
\lambda' \, \in \, \frac{1}{2^{n/2}(2\pi)^{n-1}} \, {\mathbb Z}
 = \frac{1}{2(2\pi)^{n-1}} \, \frac{\mathbb Z}{2^{n/2 - 1}} \;. 
\end{equation}
Obviously (\ref{redlquantPo}) is not the same as (\ref{lquantPo}) (though (\ref{lquantPo}) is included in
(\ref{redlquantPo})). In addition one could again use Hirzebruch signature theorem and Atiyah-Singer index theorem to gain more information and try to relax the quantization condition (\ref{redlquantPo}).
 
Up to now we have analyzed each trace product monomial (\ref{Pntraces}) independently. In cases when there are several such terms in the Lagrangian of the theory, it is only necessary that the total contribution of all topological terms in the action satisfy condition (\ref{PkFcond}). For some combinations (choice of coefficients) interference between terms may produce milder quantization conditions than those which would follow from treating each monomial separately. As a simple example illustrating this, let us specify to 7-dimensional spacetime with spin structure and assume that all topological Lagrangian terms are contained in $P_4(\R)$ which is given by
\begin{equation} \label{PnD7}
P_4(\R) =  \tr(\mathbf{R}^4) - \frac{1}{4} \left( \tr(\mathbf{R}^2) \right)^2
\end{equation}
Using formulas from Appendix \ref{app:indexthrm} it is easy to show that\footnote{The expression in 
(\ref{PnD7}) is also equal to the index of spin-$3/2$ differential operator $i D_{3/2}$ acting on 
Rarita-Schwinger field in eight dimensions.}
\begin{equation} \label{D7trint}
\frac{2}{3(4\pi)^4} \int_\mathcal{Z} P_4(\R) = 16 (\nu_+ - \nu_-) - \tau(\mathcal{Z})
 \in {\mathbb Z}
\end{equation}
By using (\ref{D7trint}) in (\ref{PkFcond}) we obtain the following quantization condition for the coupling constant corresponding to (\ref{PnD7})
\begin{equation} \label{qcondD7}
\lambda \, \in \, \frac{1}{24\, (2\pi)^3} \, {\mathbb Z}
\end{equation}
The condition (\ref{qcondD7}) is less restrictive than the one obtained by combining conditions for monomials (\ref{Pnirrtr}) and (\ref{Pnredtr}) (summed in the fixed combination (\ref{PnD7})) independently.

It is worth noting that pure gravitational Chern-Simons term (\ref{PnD7}) appears in low-energy effective string actions in some compactifications to $D=7$.  An example is 11-dimensional M-theory defined on 
$X \times M_4$, where $M_4$ is a closed compact 4-dimensional manifold, and $X$ is the actual 7-dimensional spacetime. The corresponding coupling constant is given by
\begin{equation} \label{qcondM7}
\lambda_N \, \in \, \frac{N}{192\, (2\pi)^3}
\end{equation}
where $N \in {\mathbb Z}$ when $p_1(M_4) = 0$ \cite{WittenC}.
In general, such compactifications produce other topological terms in the $D=7$ effective action, such as mixed gauge-gravitational CS terms and pure gauge CS terms, so one cannot simply compare 
(\ref{qcondM7}) with (\ref{qcondD7}). Let us assume a situation in which those other terms are either missing (for example, when $X_4$ is such that $p_1(M_4) = 0$, then mixed gauge-gravitational CS terms do not appear) or do not influence our global analysis, so that (\ref{PnD7}) is the only relevant topological term. In this case we can compare (\ref{qcondM7}) with (\ref{qcondD7}) and we see that they differ by a factor of eight. This means that such compactification scheme is consistent only when additional constraints are put on the topological numbers in (\ref{D7trint}) and (\ref{qcondM7}). There are three possibilities: (a) the Hirzebruch signature of $\mathcal{Z}$ satisfies $\tau(\mathcal{Z}) = 0 \, (\mathrm{mod} \, 8)$\footnote{For example, all manifolds which are boundaries of some closed compact manifold have vanishing Hirzebruch signature.}, (b) the winding number of the 3-form $C$ around $M_4$ satisfies $N = 0 \, (\mathrm{mod} \, 8)$, (c) some combination of properties on $X$ and $M_4$ produce jointly a factor 8.

Let us summarize the results of this subsection. First, additional (physical) requirements may reduce the configuration space of the theory, allowing for more choices for the CS coupling constants. Of course there is no guarantee that for any value of the coupling constant allowed by a given quantization condition, there will be at least one gravity theory with this coupling constant that can be consistently quantized. 
Second, since the CS coupling constant appears linearly in the entropy formula, its quantization will affect the well-definiteness of the entropy formula itself, as we be clarified in Section \ref{sec:entgcov}. 

This said, one may take the attitude of considering the CS action entirely classically (not inserted in a path integral). In this case of course there is no coupling quantization and $\lambda$ is just a free parameter.

\subsubsection{Minkowski version of the CS action}

The Minkowski version of $I_{CS}$ is not very different from the Euclidean version, due to the $i$ which, in the path integral, is present in front of both. The only difference is that the structure group $SO(2k-2,1)$ is not compact.\footnote{We recall that the $P_n$ polynomials are defined by symmetric traces of the Lie algebra generators. What changes with the Wick rotation is that antisymmetric generators of  $SO(2n-1)$, when they involve the time index, are replaced by generators that are represented by traceless symmetric matrices, see the example of $SO(1,1)\to SO(2)$ at the end of section 3.1. For instance, for the generator in the 01 plane, in the Euclidean case we have $(L^{01})_{01}=-(L^{01})_{10}$; after the Wick rotation we have $(L^{01})_{01}= (L^{01})_{10}$. However, inside $P_n$ in order to multiply the generators we need to raise the right index by means of the Minkowski metric, so that for instance we have $(L^{01})_0{}^1=-(L^{01})_1{}^0$, and the generators appear in $P_n$ effectively represented by antisymmetric matrices.} 
Only the characteristic classes (\ref{PkF}) may change (the relevant group being the maximal compact subgroup), while remaining integral. But as long as  we consider oriented base manifolds the bordism groups are the same.
Thus the conclusion is the same as in the Euclidean case, and eq. (\ref{ICSV}) with a quantized coupling is a good global representation of the CS term.

\subsection{Effects on the equations of motion}

The equation of motion does not change as a consequence of shifting from (\ref{LCS}) to (\ref{ICSV}),
because the change has a topological character, while the equation of motion is based only on local properties. In order to determine the equation of motion, $\Th$, $\J$, etc., the form (\ref{LCS}) is enough provided we eventually covariantize the final entropy formula, as we have done above.

\subsection{Global gauge invariance of the CS term}

Let $X$ be a closed manifold of dimensions $2k-1$.
If the structure group is non-Abelian, from Appendix B, for a finite (global) gauge transformation $\La$ we get
\be 
\int_X \int_0^1dt \,k\, P_k(A,F_t,F_t,\ldots,F_t) &\longrightarrow&
\int_X\Bigl[ \int_0^1dt \,k\, P_k(A,F_t,F_t,\ldots,F_t)\label{finitegauge}\\
&-& \frac{\Gamma(k)^2}{\Gamma(2k)} \, P_n(\La d\La^{-1},d\La d\La^{-1}, \ldots, d\La d\La^{-1})\Bigr]\0
\ee
the second term in the RHS is a topological term (closed but not exact). In this expression we have dropped all the total derivative terms that appear in the formulas of Appendix B. The terms we have dropped contain $A,F$ and differentials of $\La$. Is this permitted? Using this procedure it is clearly very hard to answer this question.\footnote{In the particular case of CS Lagrangian terms in $D=7$ dimensions ($n=4$) with a restriction of the topology of spacetime (after 1-point compactification) to $S^7$ in \cite{Lu:2010sj}, this procedure was used to obtain a particular quantization condition on the coupling constant corresponding to the CS term \ref{Pnirrtr}, while coupling constant corresponding to the CS term 
\ref{Pnredtr} was not quantized. However, the question is: is such topological restriction in the path integral meaningful?} If we use instead (\ref{ICSV}) it is extremely easy, the RHS is clearly invariant under any gauge transformation either infinitesimal or finite.

\section{Entropy formula and global covariance}
\label{sec:entgcov}

We have already seen that  
\be\label{SCSpt}
S_{\mathrm{CS}} &=& 2\pi \lambda n(n-1) \int_0^1 dt \int_{\mathcal{B}} P_n(\hat\epsilon,\hat\al,\hat\R_t^{n-2})
\ee
where the trace is taken over the Lie algebra of $SO(1,1)$, can be rewritten as
\be
S_{\mathrm{CS}}= 4\pi n \, \lambda \int_{\mathcal{B}}  \ome (d\ome)^{n-2}\label{SCSabel}
\ee
where we have set $\ome \equiv \hat\al^{01}_\perp$. $\ome$ is one of the two identical off-diagonal elements in the matrix representation of the connection in the normal bundle of ${\mathcal B}$ within $X$.

Now, as we have done for the action, in order to study the topological properties of the integral  in (\ref{SCSabel}) we may consider its Euclidean version. What changes is that $\lambda\to i\lambda$ and the gauge
group $SO(1,1)\to SO(2)$, while the formula to be used is (\ref{SCSpeu}). From a geometrical viewpoint
the gauge group of the normal bundle is either $SO(1,1)$ or $SO(2)$. We shall denote by $ON({\mathcal B})$ the associated principal bundle.

Formula (\ref{SCSabel}) is covariant under local coordinate transformations and local normal bundle gauge transformations. Now we would like to study the well-definiteness of this formula and its corresponding Euclidean version (\ref{SCSpeu}) as well as their response under global gauge transformations.
 
\subsection{Topological ambiguity}

Let us consider the second line of (\ref{SCSabel}) and take any $2n-2$-dimensional bounding manifold ${\mathcal W}_1$ such that $\del {\mathcal W}_1= {\mathcal B}$. Using Stokes theorem we get
\be
\int_{\mathcal{B}}  \ome (d\ome)^{n-2}= \int_{{\mathcal W}_1} (d\ome)^{n-1}\label{W1}
\ee
The same is true for any other bounding surface ${\mathcal W}_2$ such that $\del {\mathcal W}_2= {\mathcal B}$. Therefore (\ref{SCSabel}) is defined up to
\be
\int_{{\mathcal W}_1-{\mathcal W}_2} (d\ome)^{n-1}=
\int_{\mathcal Y} (d\ome)^{n-1} \label{W1-W2}
\ee
where ${\mathcal Y}$ is the closed manifold obtained by gluing ${\mathcal W}_1$ to ${\mathcal W}_2$ with reversed orientation. Thus ${\mathcal Y}$ is a closed oriented manifold of dimension $2n-2$. It is immediate to conclude that 
\be
\int_{\mathcal Y} (d\ome)^{n-1}= \int_{\mathcal Y}d(\ome (d\ome)^{n-2})=0\label{Y}
\ee
Thus there is no topological ambiguity in formula (\ref{SCSabel}). However this conclusion is not the end of the story. In fact eq.(\ref{SCSabel}), as it is written, is oversimplified.
The point is that, if we consider, as we have to, $\ome$ as a connection in a fiber bundle,
$\ome$ is in general not globally defined on the base manifold. The previous manipulations hold only if the bundle is trivial. 
\begin{itemize}
\item This is the case if the gauge group is $SO(1,1)$, because this group is contractible and any bundle with a contractible group is trivial.
\end{itemize}
To complete the analysis in this case we have to answer the question of whether there always exists a bounding
manifold for our $\mathcal{B}$. Since $\mathcal{B}$ is an oriented manifold we can use the results (\ref{cobSO})
on cobordism group. We see that we could have a problem when $D=7,11$. However, the manifold $\mathcal{B}$
can belong to a nontrivial class of $\Omega_5^{SO}$  only if it has torsion. If we exclude this case,
bounding manifolds always exist for $\mathcal{B}$. With this exclusion {\it it follows that formula (\ref{W1}) is globally well-defined.}

\vskip 1cm

Let us consider the Euclidean case next. The formula for the entropy is (\ref{SCSpt}). 
The connection $\hat\al_\perp$ and curvature $\hat\R_\perp$ take value in the Lie algebra of $SO(2)$.
They can be written in terms of $\al$ and ${\boldsymbol{\rho}}=d\al$\footnote{Once again $ {\boldsymbol{\rho}}=d\al$  is true in the total space of the bundle, it may not be true globally in the base space.} and formula (\ref{SCSp}) can be replaced by (\ref{SCSpeu}), or, more precisely, by
\be \label{SCSpeub}
S_{\mathrm{CS}} = 4\pi\,i\,\lambda\,  n \int_{\mathcal{B}} \al \,{\boldsymbol{\rho}}^{n-2} 
\ee
Again, the RHS of (\ref{SCSpeub}) is not well defined in general. We may have to replace it with
\be
S_{\mathrm{CS}} &=& 4\pi i \lambda n \int_{\mathcal W} {\boldsymbol{\rho}}^{n-1}\label{SCSabel1}
\ee
where ${\mathcal W}$ is any manifold that bounds ${\mathcal B}$. Now the integrand in (\ref{SCSabel1}) is globally defined in ${\mathcal W}$, but the integral is ambiguous unless 
\be
\int_{\mathcal Y}{\boldsymbol{\rho}}^{n-1}=0 \label{Y1}
\ee
for any closed oriented manifold of dimension $2n-2$. If this condition is satisfied then
(\ref{SCSabel1}) is well defined and can be taken as the definition of the CS entropy.

As above the passage from (\ref{SCSpeub}) to (\ref{SCSabel1}) is a nontrivial matter. The necessary condition to be satisfied is the existence of a bounding
manifold ${\mathcal W}$ for ${\mathcal B}$. As we already know this is related to the relevant cobordism group, more precisely to $\Omega_{2k-3}(BU(1),{\mathbb Z})$ (see \cite{DW}). Here $BU(1)$ is the universal classifying space of the group $U(1)$. By definition any $U(1)$ bundle can be obtained by pulling back the universal bundle $EU(1)$ over $BU(1)$ by means of a smooth map $f: {\mathcal B}\to BU(1)$. Fortunately the classifying space $BU(1)$ is very well known in various forms: ${\mathbb C}P^{\infty}$, $PU({\mathcal H})$
and in particular $K({\mathbb Z},2)$. $K({\mathbb Z},2)$ is an Eilenberg-MacLane space,
characterized by the fact that its homotopy groups vanish except for the second: 
\be
\pi_i (K({\mathbb Z},2))=0,\quad i\neq 2,\quad\quad\quad \pi_2(K({\mathbb Z},2))={\mathbb Z}
\label{homotoK}
\ee
Since $BU(1)$ has no torsion, it follows that $\Omega_{2k-3}(BU(1),{\mathbb Z})=H_{2k-3}(BU(1),{\mathbb Z})$. This means in particular that 
\be
\Omega_{2k-3}(BU(1),{\mathbb Z})=0\label{OmegaBU1}
\ee
Therefore it is always possible to replace (\ref{SCSpeub}) with (\ref{SCSabel1}). 

A sufficient condition for (\ref{Y1}) to be true is that ${\boldsymbol{\rho}}$ is globally exact, i.e. ${\boldsymbol{\rho}}=d\al$ with $\al$ globally defined on ${\mathcal B}$. This is possible if the $U(1)$ bundle is trivial. In this case the bundle can be trivially extended to any ${\cal W}$. Now, when are we sure that the $ON({\mathcal B})$ bundle is trivial? The $U(1)$ or line bundles ($SO(2)$ or circle bundles) are classified by the cohomology group $H^2({\mathcal B},{\mathbb Z})$. In the case
${\mathcal B}$ is an odd dimensional sphere (or any homeomorphic manifold) this group is trivial, thus in this case the above requirement is satisfied. Other interesting horizon topologies in $D \ge 5$ most frequently discussed in the literature\footnote{General classification of allowed horizon topologies for black holes in $D \ge 7$ is largely unknown (for recent reviews see \cite{Galloway,HoIsh}). In General Relativity there are numerical solutions for asymptotically flat black holes in $D = 7$ with $S^1 \times S^{D-3}$ (``black rings'') \cite{KlKuRa} and $S^2 \times S^{D-4}$ (``generalized black rings") \cite{KlKuRaRo} horizon topologies. In \cite{Schwartz} asymptotically flat black hole spacetimes with horizon topologies 
$S^{k} \times S^{D-k-2}$ were explicitly constructed, however it is not known are these spacetimes solutions in any (generalized) gravity theory.} are  ``generalized black rings" with $S^{k} \times S^{D-k-2}$, 
$1 \le k \le D-3$ topology. For them the second cohomology group is trivial, except in the case where $k=2$. In general, if ${\mathcal B}$ is an odd dimensional torus $T^{2n-3}$, or contains a 2-sphere or an $n$-torus ($n\geq 2$) as a factor then  $H^2({\mathcal B},{\mathbb Z})\neq 0$. In this case the possibility of a nontrivial normal bundle would have to be considered.

\begin{itemize}
\item In the case the above triviality requirement (\ref{Y1}) is satisfied, 
(\ref{SCSabel1}) {\it is a good definition for the entropy from a CS term in the Euclidean case}, since it is free of topological ambiguities. If the normal bundle is trivial this definition coincides with 
(\ref{SCSpeub}).
\end{itemize}

\subsection{Nontrivial horizon geometries}
\label{ssec:nthorg}

We have seen that the triviality requirement (\ref{Y1}) may not always be met, in which case formula (\ref{SCSabel1}) is not free of topological ambiguities. Even when (\ref{SCSabel1}) is unambiguous 
it may be rather unpractical for the purpose of an explicit calculation.
It is preferable, whenever possible, to use the local formula (\ref{SCSpeu}) or (\ref{SCSpeub}). However, as we shall see in a moment, global gauge transformations may be the origin of ambiguities for these formulas too. This problem requires a slight extension of the definition of entropy, which is 
suggested by a path integral formulation.
Following \cite{BY} we can start from the canonical partition function of a generic statistical system $Z(\beta)= \int dE \,\nu(E) \, e^{\beta E}$, where $\nu(E)$ is the density of states with energy $E$. 
This can be rewritten as 
\be
Z(\beta)= \int dE\, e^{-I(E)}\label{Zbeta}
\ee
where $I(E)= \beta E-S(E)$, with the entropy function $S(E)$ defined by $S(E)=\log \nu(E)$. $I(E)$ can be interpreted as the analog of the Euclidean  action. The precise identification poses several problems, studied for instance in \cite{BY}. Here we are interested
only in the generic linear relation between Euclidean action and entropy function. Simply this means that
if we compute the entropy via a path integral, it will appear as the logarithm of a certain expression. If the latter is defined up to $2\pi i {\mathbb Z}$, the entropy is well defined. Therefore
the entropy function corresponding to a CS term in the Euclidean version can have ambiguities analogous 
to the CS action, ambiguities which may be resolved due to the fact entropy is at the
exponent and, thanks to Wick rotation, is multiplied by $i$. Thus, these 
ambiguities are harmless if, when suitably normalized, they are of the form $2\pi\,i\, k$ with $k\in {\mathbb Z}$, that is, if
 \be \label{ambig1}
\Delta S_{CS}^{(E)} \, \in \, 2\pi i {\mathbb Z}
\ee
In particular formula (\ref{Y1}) is replaced by the milder condition
\be \label{Y2}
\lambda \int_{\mathcal Y}{\boldsymbol{\rho}}^{n-1} \, \in \, \frac{\mathbb Z}{2n}
\ee
Thus the statement at the end of the previous subsection becomes
\begin{itemize}
\item In the case the requirement (\ref{Y2}) is satisfied, 
(\ref{SCSabel1}) {\it is a good definition for the entropy from a CS term in the Euclidean case} as it is free of topological ambiguities.  
\end{itemize}

In conclusion we have to ask ourselves when (\ref{Y2}) is true. For a $U(1)$ bundle with curvature
$\mathbf{F}$ on a $2m$-dimensional base space $\mathcal{Y}$ we know in general that
\begin{equation} \label{chnum1m}
\frac{1}{(2\pi)^m} \int_\mathcal{Y} \mathbf{F}^m = C_1^m \, \in \, {\mathbb Z}
\end{equation}
where $C_1^m$ is one of the Chern numbers of the line bundle (see Appendix \ref{sapp:gCSent}). We can use this in our case where $\mathbf{F} = \boldsymbol{\rho}$ and $m = n-1$.

As already noted above, when at least one of cohomology groups 
$H^{2}(\mathcal{Y})$ or $H^{2(n-1)}(\mathcal{Y})$ is trivial, then obviously $C_1^{n-1}(E) = 0$, which means that (\ref{Y1}) will be satisfied. However, this condition is not necessary.

If the line bundle is such that $C_1^{n-1} \ne 0$, we must analyze the consequences of (\ref{chnum1m}) on 
condition (\ref{Y2}). If $C_1^{n-1}$ assumes integral values (unrestricted topology), then by combining (\ref{chnum1m}) and (\ref{Y2}) the following condition on $\lambda$ follows
\begin{equation} \label{htopqc}
\lambda = \frac{1}{2(2\pi)^{n-1}} \, \frac{\mathbb Z}{n}
\end{equation}
Given these conditions we have to compare (\ref{htopqc}), obtained by requiring an unambiguous definition of the entropy of (Euclidean) black holes, with the conditions
(\ref{lquantPo}), (\ref{lquantHi}) or (\ref{qcondD7}) obtained in Sec. \ref{sssec:euCSact} by requiring an unambiguous definition of the original path integral (the second and third conditions come from different restrictions on the configuration space). We see that (\ref{htopqc}) is different, but that 
while (\ref{lquantPo}) is encompassed by (\ref{htopqc}) (in the sense that the couplings of type (\ref{lquantPo}) certainly satisfy (\ref{Y2})), conditions (\ref{lquantHi}) and (\ref{qcondD7}) may not (depending on the actual value of the coupling). However it should be kept in mind that there may be topological restrictions on the bifurcation horizon $\mathcal{B}$, which in principle may induce restrictions on the possible values of Chern numbers $C_1^{n-1} \ne 0$. This in turn may produce a condition on $\lambda$ less restrictive than (\ref{htopqc}), and this might conspire to include the (\ref{lquantHi}) case. In other words {\it validity of the
(\ref{Y2}) and its agreement with the coupling quantization conditions have to be checked carefully case by case, taking into account all possible restrictions of the topology and, more in general, of the configuration space of the theory}. 

Let us finish this analysis with the following observation. In the case $n=2$, which means in the context of 3-dimensional gravity, condition (\ref{htopqc}) \emph{exactly} matches the choices for $\lambda$ emphasized by Witten in \cite{Witten}. There the condition came from requiring holomorphic factorization, and, interestingly, on some consistency requirements in the Ramond-Ramond black hole sector. One may consider the possibility that condition (\ref{htopqc}) has some role for quantum gravity in higher dimensions.

\subsection{Covariance under global gauge transformations}

In the Minkowski case, as we have seen, the CS entropy formula is always given by (\ref{SCSabel}). The problem is to prove invariance of this formula under global gauge transformations. 
In the Minkowski case the normal bundle group is $SO(1,1)$. This group is contractible, so also the group of gauge transformations is contractible and there is nothing beyond the local gauge transformations: the problem of global transformations simply does not exist.

In the Euclidean case we have more possibilities.
The global formula (\ref{SCSabel1}) is evidently invariant under all gauge transformations (whether local or global). However, as we have noticed, this formula may be unpractical.
It is important to know when we can avail ourselves of the local formula (\ref{SCSpeub}). If the
relevant $U(1)$ gauge bundle is nontrivial there is nothing we can do but use (\ref{SCSabel1}), and in this case the problem of global gauge transformations is irrelevant. Otherwise the natural candidate is 
(\ref{SCSpeub}), except that we have to prove its invariance under global gauge transformations.

In the Euclidean case (\ref{SCSpeub})  expresses a $2n-3$ form integrated over the $2n-3$ dimensional surface ${\mathcal{B}}$. However it can also be read as 
an Abelian CS term, determined by the connection $\omega$ valued in the Lie algebra of the group  
$U(1)$. On the basis of formula (\ref{Pnabelian}) in Appendix B, the term corresponding to the second one in the RHS of (\ref{Pn}) becomes (using (\ref{SCSpeu},\ref{gaugetrso2}) and (\ref{gaugeso2}))
\be\label{global1} 
4\pi i \lambda n \int_{\cal B} d\left(\al {\boldsymbol {\rho}}^{n-3} dg\right)=4\pi i \lambda n \int_{\cal B} d\left({\boldsymbol {\rho}}^{n-2}\, g\right)
\ee
This of course vanishes if $g$ is single valued. But it may not be so if $g$ is multivalued. To appreciate
this point it is instructive to consider first the $n=2$ case (we include for comparison also the Minkowski case)

When $n=2$ CS entropy is
\be \label{n=2}
S_{\mathrm{CS}}= 8\pi\, \lambda \int_{S^1}  \ome \;\;, \qquad \quad
S_{\mathrm{CS}} = 8\pi\,i\,\lambda\, \int_{S^1} \al 
\ee
in the Minkowski and Euclidean case, respectively. Formula (\ref{Pnabelian}) becomes  
\be 
\int_{S^1} \ome   \rightarrow \int_{S^1} \ome - \int_{S^1} d\theta \,\del_\theta f(\theta),\quad\quad\quad
\int_{S^1} \al   \rightarrow \int_{S^1} \al - \int_{S^1} d\theta \,\del_\theta g(\theta)
\label{casen=2}
\ee
It is evident from these definitions that we are calculating the {\it holonomy} of the relevant connections.

\begin{itemize}
\item Minkowski case. In this case $f$ is periodic in $\theta$ and single-valued. It follows that
\be
\int_0^{2\pi} d\theta \partial_\theta f(\theta)=0\0
\ee
{\it This confirms that in this case there is no problem with global gauge transformations.}

\item Euclidean case. In this case the gauge group is $SO(2)$. Then the most general global transformation  takes the form (\ref{gaugetrso2})
\begin{equation}
\Lambda =  \left( \begin{matrix} \cos g & \sin g \\ -\sin g & \cos g \end{matrix} \right)\0
\end{equation}
Periodicity of $\Lambda$ implies  $g(\theta + 2\pi) = g(\theta) + 2\pi k$, with integer $k$. In this case (\ref{gaugeso2}) becomes 
\be 
\al\to \al +\del_\theta g\, d\theta\0
\ee
Integration in (\ref{casen=2}) then gives
\be \label{globalso2}
\int_0^{2\pi} d\theta \, g'(\theta) = g(2\pi) - g(0) = 2\pi \, k
\ee
Thus, from (\ref{casen=2}), the ambiguity of the Euclidean entropy (\ref{n=2}) is 
\be\label{ambig}
\Delta S_{CS}^{(E)} = \lambda (4 \pi)^2 i\, k \;\;, \qquad k \in {\mathbb Z}
\ee
Using quantization condition (\ref{lquantPo}), which for $n=2$ reads
\begin{equation} \label{lqPon2}
\lambda \in \frac{1}{4\pi} {\mathbb Z} \;\;,
\end{equation}
in (\ref{ambig}) we obtain
\be\label{ambigc1}
\Delta S_{CS}^{(E)} \, \in \, 4 \pi\, i\,  {\mathbb Z} \, \subset \, 2 \pi\, i\,  {\mathbb Z}
\ee 
which falls among the (\ref{ambig1}) cases.

If, instead, we use the weaker quantization condition (\ref{lquantHi}) corresponding to non-singular geometrical configurations in gravity, which for $n=2$ reads
\begin{equation} \label{lqhalf}
\lambda \in \frac{1}{12\pi} {\mathbb Z} \;\;,
\end{equation}
the ambiguity becomes
\be\label{ambig2}
\Delta S_{CS}^{(E)} \, \in \, \frac 43 \pi\, i\, {\mathbb Z} \, \not\subseteq \, 2 \pi\, i\,  {\mathbb Z}
\ee
Similarly, for the quantization condition (\ref{lquantASn2})
\begin{equation} \label{lqASn2}
\lambda \in \frac{1}{48\pi} {\mathbb Z} \;\;,
\end{equation}
corresponding to configurations allowing spin structure, we obtain
\be\label{ambig3}
\Delta S_{CS}^{(E)} \, \in \, \pi\, i\, \frac{\mathbb Z}{3} \, \not\subseteq \, 2 \pi\, i\,  {\mathbb Z}
\ee
{\it Therefore in these two cases the entropy formula (\ref{SCSpeu}) may be ambiguous (depending on the specific value of the coupling) .}

We could turn the question around and ask what are the values of $\lambda$ for which the ambiguity 
(\ref{ambig}) is certainly in accord with (\ref{ambig1}). The answer is
\begin{equation} \label{lqcsbh}
\lambda \, \in \, \frac{1}{8\pi} \, {\mathbb Z}
\end{equation}
which is the same result as (\ref{htopqc}) obtained from a more general analysis. 
\end{itemize}

The ambiguities (\ref{ambig1}) and (\ref{ambig2}) are due to the fact that the function $g(\theta)$ is multi-valued. It represents a map from the circle $S^1$ (the bifurcation surface) to the circle representing the space $SO(2)$. We can now consider generic $n$, with ambiguity given by (\ref{global1}). The integrand represent a map from ${\mathcal B}$ to $S^1$. These maps fall into classes
classified by the cohomotopy group $\pi^1({\mathcal B})$. The latter is isomorphic to the cohomology
group $H^1({\mathcal B},{\mathbb Z})$. For instance, if ${\mathcal B}$ is $S^{2n-3}$, then the corresponding cohomology group is trivial, except for $n=2$. In any case, provided the normalization is the correct one the integral in (\ref{global1}) yields an integer multiplied by $2\pi$, and this would fall again among the cases (\ref{ambig1}). However this has to be verified case by case
because the normalization of the coupling constant may not coincide, in general, with the
normalization of the cohomology classes in $H^1({\mathcal B},{\mathbb Z})$.

\section{Conclusions}

In this paper we have analyzed two related issues. The first is the global invariance of the gravitational Chern-Simons action term. This problem is well-known and well analyzed in the simplest case ($n=2$, corresponding to $D=3$). Here we have addressed the same problem in general as far as it is
possible to give general answers, and for the lowest $n$ cases when this is not possible. We have 
identified the general topological conditions under which the CS term is globally covariant (though this may require a modification of the CS formula, from (\ref{LCS}) or (\ref{LCSa}) to (\ref{PkR})). As is well-known this has also an important consequence: unless the theory we are analyzing is considered from a strictly classical point of view, the CS coupling must be quantized. We have also seen that these quantization conditions may be to some extent relaxed by restricting the allowed geometries in the theory,
which is required in the gravitational case. As we have seen this may have direct consequences for the CS entropy formula. 

The second issue we have treated is the global covariance of the CS entropy formula. The analysis for the latter is somewhat similar to that for the CS term, because the CS entropy formula can be read itself as an Abelian CS term. We have verified that the condition for global covariance are generally satisfied in the Minkowski version of the theory, but the situation is more complicated in the Euclidean version. In the latter case again we can generally satisfy the conditions for global covariance, although at the price of 
shifting to formula (\ref{SCSabel1}) from (\ref{SCSpeu}). However there may be nasty cases, connected to
the relaxed CS coupling constant, in which this is not possible and even (\ref{SCSpeu}) is ambiguous. Finally, we have studied the invariance under global gauge transformations in cases in which formula (\ref{SCSabel1}) is topologically well defined. We have found that, again, relaxed CS couplings  
may forbid in some cases the use of this formula.

A final comment concerns the validity of our result in the case of AdS black holes. In this regard two issues should be kept in mind: the derivation of the entropy formula and the coupling quantization of the CS terms. As for the entropy formula, there seems to be no obstruction to the validity of our derivation in 
\cite{BCDPS1} for the same reason as for Wald's formula. As for the second issue, the presence of a boundary  changes the discussion with respect to the Minkowski case. In particular in order to pass from (\ref{ICSX}) to (\ref{ICSV}) we must use another version of cobordism, the relative cobordism or cobordism for manifolds with boundary, and, specifically, with the presence of some structure both on the bulk and the boundary. To our best knowledge this problem has not been dealt with in the literature and we would like to return to it in a future occasion.

\acknowledgments 

One of us (L.B.) would like to thank the Physics Department, Univ.\ of Zagreb, for hospitality and financial support during his visits there. 
The work of L.B. has been supported in part by the MIUR-PRIN contract 2009-KHZKRX. M.C., P.D.P., 
S.P.\ and I.S.\ would like to thank SISSA for hospitality and financial support during visits there and would also like to acknowledge support by the Croatian Ministry of Science, Education and Sport under the contract no.\ 119-0982930-1016. Visits have been financially supported in part by the University of Zagreb Development Fund.

\section*{Appendix}
\appendix

\section{A local coboundary formula}

In order to find the relation between (\ref{LCS}) and (\ref{LCSa}) one could follow \cite{BZ}. One introduces the symbol $H=\log E$ and the interpolating connection
\be
\Gam_s= e^{-s H} d e^{s H}+  e^{-s H} \Gam e^{s H}\label{Gammas}
\ee
We see that $\Gam_0=\Gam$ and $\Gam_1=\hat\al$. We have also $\R_s = e^{-s H}\R e^{s H}$. 
We introduce next $\Gam_{s,t} = t\Gam_s$ and $\R_{s,t}= d\Gam_{s,t} +\Gam_{s,t} \Gam_{s,t}$. We notice that
\be
\frac d{ds} \Gam_s= e^{-s H} {D_\Gam H} e^{s H}\equiv {\cal H}_s \label{dsGam}
\ee
and
\be
\frac d{ds} \R_{s,t}= D_{\Gam_{s,t}} \left(t{\cal H}_s\right)\label{dsR}
\ee

Now we start from the identity
\be
\CS^{(n)}_{\mathrm{CS}}(\hat\al)-\CS^{(n)}_{\mathrm{CS}}(\Gam)=n \int_0^1 ds \frac d{ds}
\int_0^1 dt \, P_n\left(\Gam_s,\R_{s,t}^{n-1}\right)\label{CS-CS}
\ee
and perform the derivative with respect to $s$ explicitly. Using (\ref{dsGam}) and (\ref{dsR}) and the by now familiar tricks it is easy to show that
\be
\CS^{(n)}_{\mathrm{CS}}(\hat\al)-\CS^{(n)}_{\mathrm{CS}}(\Gam)= d\Big( n\, P_n(H,\R^{n-1})-
n(n-1) \int_0^1ds\int_0^1 dt\ P_n(\Gam_{s,t},{\cal H}_s, \R_{s,t}^{n-2})\Big)
\label{CS-CSd}
\ee
However all this can have only a local meaning. For instance the quantity $P_n(H,\R^{n-1})$ transform in a very nonlinear way under general coordinate or Lorentz transformations. Since $H$ has one flat and one world index,
in  $P_n(H,\R^{n-1})$ one flat index is saturated with one world index, which badly breaks 
covariance. In fact the difference between the two CS terms in (\ref{CS-CSd}) contain a non-exact (topological)
term. However (\ref{CS-CSd}) tells us that (\ref{LCS}) and (\ref{LCSa}) are related by local irrelevant terms.

\section{Global gauge transformations} 

Let us consider a gauge theory with connection $A$ and gauge group $G$. A finite gauge transformation takes the form
\be
A\rightarrow \Lambda^{-1} (d+A)\Lambda\label{finite}
\ee
with $\Lambda$ valued in $G$. The problem we want to solve is to find the transformation of a Chern-Simons term
under a finite gauge transformation. We can write
\be
A\rightarrow \La^{-1}(A-\La d\La^{-1})\La ,\quad\quad F_t
\rightarrow \La^{-1}\bigl(F_t+(t-t^2)D (\La d\La^{-1})\bigr) \La\label{finite2}
\ee
where $F_t=tdA +t^2A^2$.
It is not possible to use the same tricks as for infinitesimal transformations. With a direct approach
one finds
\be
\int_0^1dt P_2(A,F_t) \rightarrow \int_0^1dt P_2(A,F_t) +\frac 16 P_2(\La d\La^{-1}, \La d\La^{-1}\La d\La^{-1})-
\frac 12 d\,P_2(A,\La d\La^{-1})\label{P2}
\ee
The term $P_2(\La d\La^{-1}, \La d\La^{-1}\La d\La^{-1})$ is a topological term (it is closed but not exact as a form).
Similarly we have
\be
&&\int_0^1dt P_3(A,F_t,F_t) \rightarrow \int_0^1dt P_2(A,F_t,F_t) -\frac 1{60} P_3(\La d\La^{-1}, \La d\La^{-1}\La d\La^{-1},\La d\La^{-1}\La d\La^{-1})\0\\
&& -d\Bigl[ \frac 16 \,P_3(A,\La d\La^{-1},d\La d\La^{-1})+ \frac 13 P_3(A,F,\La d\La^{-1}) +\frac 16 P_3(A,\La d\La^{-1},[A,\La d\La^{-1}])\Bigr]\label{P3}
\ee
Again the second term in the RHS is the topological term. The other terms are exact. In general one can prove that
\be
\int_0^1dt P_n(A,F_t,F_t,\ldots,F_t)&\rightarrow& \int_0^1dt\int_0^1dt P_n(A,F_t,F_t,\ldots,F_t)\label{Pn}\\
&&- \frac{\Gamma(n)^2}{\Gamma(2n)} \, P_n(\La d\La^{-1},d\La d\La^{-1}, \ldots, d\La d\La^{-1})\0\\
&&-d\Bigl[ \frac{\Gamma(n)^2}{\Gamma(2n-1)}\, P_n(A,\La d\La^{-1},d\La  d\La^{-1}, \ldots, d\La d\La^{-1})+\ldots\Bigr]\0
\ee
dots denote exact terms not yet calculated 

When the group is abelian things simplify considerably. It is easy to prove that
\be
\int_0^1dt P_n(A,F_t,F_t,\ldots,F_t)= P_n(A,dA,\ldots,d A) -dP_n(A,dA,\ldots,dA,\La  d\La^{-1})\label{Pnabelian}
\ee
The reason why the other terms vanish is that, when the group is abelian, $d\La  d\La=0$, etc.

\section{On the connection ${\omega}$ in gravitational CS entropy term}
\label{app:eugCSent}

Here we want to analyze in more detail the properties of the connection $\omega$, discussed also in section 3.1, both in Lorentzian and Riemannian contexts.

We have shown in \cite{BCDPS1} that a CS black hole entropy term can be written as a CS term by
using 
\begin{equation} \label{omeganl}
\omega_\mu = - q_\mu^\nu \, n_\rho \nabla_\nu \ell^\rho
\end{equation}
The spacetime here is Lorentzian, and $n$ and $\ell$ are null-vectors orthogonal to the bifurcation surface
$\mathcal{B}$. Null-vectors are normalized such that
\begin{equation} \label{nlnorm}
n^2 = 0 = \ell^2 \;\;, \qquad n \cdot \ell = -1
\end{equation}
Obviously there is a freedom in choosing a null-basis, which is described by a local transformation
\begin{equation} \label{gcmin}
\ell \to e^f \ell \;\;, \qquad n \to e^{-f} n
\end{equation}
where $f(x)$ is any smooth function. Under (\ref{gcmin}) the 1-form $\boldsymbol{\omega}$ transforms as
\begin{equation} \label{gtomega}
\boldsymbol{\omega} \to \boldsymbol{\omega} + d\,f
\end{equation}

Instead of using the null-vectors $n$ and $\ell$ we can pass to a couple of orthonormal vectors $m^{(0)}$ and $m^{(1)}$ which satisfy
\begin{equation} \label{orthomin}
m^{(a)} \cdot m^{(b)} = \eta^{ab} \;\;, \qquad a,b = 0,1
\end{equation}
where $\eta^{ab}$ is 2-dimensional Minkowski metric (in this Appendix $a,b,c$ denote flat Minkowski indices. The relation between the two bases is given by
\begin{equation} \label{mAtonl}
m^{(0)} = \frac{1}{\sqrt{2}} (\ell + n) \;\;, \qquad m^{(1)} = \frac{1}{\sqrt{2}} (\ell - n)
\end{equation}
The inverse relation is 
\begin{equation}
\ell = \frac{1}{\sqrt{2}} (m^{(0)} + m^{(1)}) \;\;, \qquad n = \frac{1}{\sqrt{2}} (m^{(0)} - m^{(1)})
\end{equation}
We can now write (\ref{omeganl}) as
\begin{equation} \label{omegam01}
{\boldsymbol \omega}\equiv\omega_\mu^{(10)} = q_\mu^\nu \, m^{(1)}_\rho \nabla_\nu m^{(0)\rho}
 = - \frac{1}{2} q_\mu^\nu \, \varepsilon_{ab} \,  m^{(a)}_\rho \nabla_\nu m^{(b)\rho}
\end{equation}
where $\varepsilon_{ab}$ is the 2-dimensional totally antisymmetric tensor with $\varepsilon_{01} = 1$.
The transformation (\ref{gcmin}) in this basis is
\begin{equation} \label{m01gauge}
m^{(a)} \to O^a{}_b \, m^{(b)} \;\;, \qquad \mbox{where} \qquad O =
 \left(\begin{matrix} \cosh f & \sinh f \\ \sinh f & \cosh f \end{matrix} \right) \in SO(1,1)
\end{equation}
which is a group of (pseudo)rotations of normal frames, represented in the fundamental two-dimensional representation. This is the expected result (see section (3.1)).

Using (\ref{m01gauge}) in (\ref{omegam01}) we can check that $\boldsymbol{\omega}$ transforms
as in (\ref{gtomega})
\begin{equation} \label{gtomegam}
\boldsymbol{\omega}
 \to \boldsymbol{\omega} - \frac{1}{2} \varepsilon_{ab} \, O^a{}_c \, \eta^{ce} d \, O^b{}_e 
  = \boldsymbol{\omega} + O^1{}_a \, \eta^{ab} d \, O^0{}_b
 = \boldsymbol{\omega} + d\,f
\end{equation}

As already pointed out $\boldsymbol{\omega}$ can be viewed as a connection in a principal
bundle with base space $\mathcal{B}$. From (\ref{m01gauge}) it is obvious that the structure group is 
$SO(1,1)$. The $m^{(a)}$'s are in the fundamental 2-dimensional representation of it. Here we have to 
explain an apparent paradox: (\ref{m01gauge}) describes a two-dimensional fiber, while (\ref{gtomega}) and (\ref{gtomegam})
suggests a one-dimensional fiber. In fact the latter refers to the transformation of a 
connection in the principal $SO(1,1)$ bundle whose fiber is in fact one dimensional but takes values in the Lie algebra of the gauge group, and the Lie algebra of $SO(1,1)$ is represented by 
a $2\times 2$ off-diagonal matrix with two identical non-zero elements. This is evident from (\ref{omegam01}) where 
only one of the two elements appears, the other is identical, with label $^{(01)}$. On the other hand (\ref{m01gauge}) refers to the corresponding transformation in the associated vector bundle defined by the two-dimensional fundamental representation of $SO(1,1)$.  
 
\bigskip

Let us now repeat this in the Riemannian spacetime. Here we can use the analogue of (\ref{omegam01})
\begin{equation} \label{omegaEm01}
\omega_\mu = q_\mu^\nu \, m^{(1)}_\rho \nabla_\nu m^{(0)\rho}
\end{equation}
where now orthonormal vectors, instead of (\ref{orthomin}), satisfy
\begin{equation} \label{orthominE}
m^{(a)} \cdot m^{(b)} = \delta^{ab} \;\;, \qquad a,b = 0,1
\end{equation}
The gauge group of frame rotations is now, of course, $SO(2)$, represented on $m^{(a)}$ with standard 
2-dimensional representation
\begin{equation} \label{gaugeEm}
m^{(a)} \to O^a_{\;\;b} \, m^{(b)} \;\;, \qquad \mbox{where} \qquad 
O = \left(\begin{matrix} \cos g & \sin g \\ -\sin g & \cos g \end{matrix} \right) \in SO(2)
\end{equation}
where $g$ is an arbitrary real function, under which the connection (\ref{omegaEm01}) transforms as
(adopting the same symbols as in the text)

Now, here there are no real null-vectors, but we can still write an analogue of (\ref{mAtonl}) by formally
using a complex vector $\xi$ defined with
\begin{equation}
\xi = \frac{1}{\sqrt{2}} (m^{(0)} + i \, m^{(1)}) \quad \Longrightarrow \quad 
\xi^* = \frac{1}{\sqrt{2}} (m^{(0)} - i \, m^{(1)})
\end{equation}
It follows that
\begin{equation}
\xi \cdot \xi = 0 \;\;, \qquad \xi \cdot \xi^* = 1
\end{equation}
Using $\xi$ we can write the Euclidean connection (\ref{omegaEm01}) as
\begin{equation} \label{omegaExi}
\omega_\mu= i \,q_\mu^\nu \, \xi^*_\rho \nabla_\nu \xi^\rho
\end{equation}
The gauge transformation (\ref{gaugeEm}) acts on $\xi$ as
\begin{equation} \label{gtxiE}
\xi \to U \xi \quad \Longrightarrow \quad \xi^* \to U^* \xi^* \;\;, \qquad U = e^{-ig} \, \in \, U(1)
\end{equation}
In so doing the have mapped an $SO(2)$ geometry into an equivalent $U(1)$ geometry. The connection
transforms again as 
\begin{equation} \label{omEgauge}
{\boldsymbol \alpha}\to {\boldsymbol \alpha} + dg,\quad\quad {\rm or}\quad\quad {\boldsymbol \beta}\to  {\boldsymbol \beta} -i dg
\end{equation}
where the notation ${\boldsymbol \beta}=-i{\boldsymbol \alpha}$ is the usual one for $U(1)$ connections. The transformation (\ref{gtxiE}) represents the $U(1)$ 
action over a one-dimensional complex fiber. The associated bundle is now a complex line bundle.
In this way (i) we have transformed the relevant bundle into a \emph{complex} one, which will allow us to use Chern classes, (ii) the normalization of the connection $\boldsymbol{\alpha}$ is the correct one to calculate Chern numbers.

\section{Characteristic classes and index theorems}
\label{app:indexthrm}

In our paper we repeatedly come across integrals of the type
\be \label{topinteg}
 \int_{\cal Z} \left( \tr(\mathbf{F}^{k_1}) \right)^{l_1} \left( \tr(\mathbf{F}^{k_2}) \right)^{l_2} \cdots
  \;\; , \qquad \sum_{r=1} k_r \, l_r = m
\ee
where $\mathbf{F}$ is a 2-form curvature in some vector bundle $E$ with fiber dimension $d_F$ associated to principle bundle with a structure group $G$ and with base space ${\cal Z}$, which is a closed oriented manifold of dimension $2m$, $m \in {\mathbb N}$. For definiteness we shall assume that $\tr$ refers to the trace in the fundamental group of $G$. Such integrals are topological quantities taking discrete values. In this Appendix we collect well-known mathematical results about them. A presentation of this material for physicists can be found for instance in \cite{AlvarezGaume:1984dr,Eguchi:1980jx,Nakahara:2003nw}.

For simplicity, as an example of (\ref{topinteg}), we will concentrate on the irreducible term
\be \label{topintegir}
 \int_{\cal Z} \tr(\mathbf{F}^m) 
\ee

\subsection{CS action term}

When analyzing the allowed values for CS coupling constant $\lambda$ in Sec. \ref{sssec:euCSact}, we were faced with (\ref{topinteg}), with $m=n \in 2 {\mathbb N}$, $G = SO(2n -1,1)$ (or $SO(2n)$). This is a real bundle for which we can use Pontryagin numbers to obtain info on (\ref{topinteg}). The total Pontryagin class is defined via
\begin{equation} \label{pontclass}
p(E) \equiv \det \left( 1 + \frac{\mathbf{F}}{2\pi} \right) = 1 + p_1(\mathbf{F}) + p_2(\mathbf{F}) + \ldots
\end{equation}
where $p_j(\mathbf{F})$, the $j$-th Pontryagin class, is a $4j$-form defining the cohomology class $p_j(E) \in H^{4j}({\cal Z})$. They are independent of the choice of $\mathbf{F}$). From (\ref{pontclass}) it follows that 
$p_j(\mathbf{F})$ can be expressed as linear combinations of products of traces of powers of $\mathbf{F}$
with rational coefficients. This relation can also be inverted. In particular, for irreducible trace, such as one appearing in (\ref{topintegir}), one has
\begin{equation} \label{trFpjk}
\frac{1}{2(2\pi)^{2j}} \, \tr(\mathbf{F}^{2j}) =
 \sum_{k_1,\ldots,k_j = 0}^j a_{k_1 \cdots k_j}^{(j)} \prod_{r=1}^j (p_r(\mathbf{F}))^{k_r} \;\;, \qquad
 \sum_{r=1}^j r \, k_r = j  
\end{equation}
It can be shown that coefficients $a_{k_1 \cdots k_j}^{(j)}$ are \emph{integers}
\begin{equation} \label{coeffint}
a_{k_1 \cdots k_j}^{(j)} \in {\mathbb Z}
\end{equation}
In particular, the coefficient of $p_1^j$ is $a_{j0 \cdots 0}^{(j)} = (-1)^j$, and the coefficient of $p_j$ is $a_{0 \cdots 01}^{(j)} = -j$, for all $j$. Unfortunately there is no simple explicit formula for the generic coefficient $a_{k_1 \cdots k_j}^{(j)}$. Let us write first few terms explicitly
\begin{eqnarray}
\frac{1}{2(2\pi)^{2}} \, \tr(\mathbf{F}^2) &=& - p_1 
 \nonumber \\
\frac{1}{2(2\pi)^{4}} \, \tr(\mathbf{F}^4) &=& p_1^2 - 2 \, p_2
 \nonumber \\
\frac{1}{2(2\pi)^{6}} \, \tr(\mathbf{F}^6) &=& - p_1^3 + 3 \, p_1 \, p_2 - 3 \, p_3
 \nonumber \\
\frac{1}{2(2\pi)^{8}} \, \tr(\mathbf{F}^8) &=& p_1^4 - 4 \, p_1^2 \, p_2 + 4 \, p_1 \, p_3 + 2 \, p_2^2 - 4 \, p_4
 \nonumber \\
&\vdots&
\label{Pontexp}
\end{eqnarray}
Using (\ref{trFpjk}) one can obtain expressions for the case of reducible traces.

What is important for our purposes is that integrals of $p_j^k$ over $4jk$-dimensional closed oriented submanifolds $\mathcal{S}$ of $\mathcal{Z}$ are \emph{integers},
\begin{equation} \label{intpjk}
\int_\mathcal{S} (p_j(\mathbf{F}))^{k} \in {\mathbb Z}
\end{equation}
If we integrate over the entire base manifold $\mathcal{Z}$ we obtain the so called Pontryagin numbers 
$P_j^k(E)$ of the bundle $E$
\begin{equation} \label{Pontnum}
P_j^k(E) \equiv \int_\mathcal{Z} (p_j(\mathbf{F}))^{k} \in {\mathbb Z}
\end{equation}
By using (\ref{trFpjk}), (\ref{coeffint}) and (\ref{Pontnum}) we obtain that
\be \label{inttrirZ}
\frac{1}{2(2\pi)^n} \int_{\cal Z} \tr(\mathbf{F}^n) \in {\mathbb Z}
\ee
In the same way one can show that similar results can be obtained for integrals of reducible traces (the coefficient on the right hand side is then product of individual coefficients $1/(2(2\pi)^{2j}$ of the each trace factor).

\bigskip

In the previous analysis we have referred to a generic vector bundle $E$ with gauge group $SO(2n-1,1)$ (or $SO(2n)$) corresponding to a standard gauge theory. However, we are interested in a theory of gravity. In classical gravity the principal bundle is the orthogonal bundle and what we called $\mathbf{F}$ is in fact the 2-form Riemann curvature $\mathbf{R}$, which (together with corresponding connection) is obtainable from a nonsingular (i.e., invertible) metric tensor (or vielbein). If we assume that in the path integral we should take into account only such configurations which are classically well-motivated, we get a constraint which we may use to obtain additional information. Let us assume this from now on.

In the mathematical language, the above means that $E$ is the tangent bundle $T{\cal Z}$, with corresponding Pontryagin classes usually denoted as $p_j({\cal Z})$. We can use the Hirzebruch signature theorem which, adapted to our case, says
\begin{equation} \label{Hbsigth}
\int_{\cal Z} L({\cal Z}) = \tau({\cal Z}) \in {\mathbb Z}
\end{equation}
where $\tau({\cal Z})$ is an index called the Hirzebruch signature. $L({\cal Z})$ is the Hirzebruch 
$L$-polynomial, which can be written as polynomial in the $p_j({\cal Z})$'s with rational coefficients. There is no general closed form expression, but, if we write $L({\cal Z})$ as
\begin{equation}
L({\cal Z}) = \sum_{j=0}^{n/2} L_j({\cal Z}),\0
\end{equation}
the first few terms are given by
\begin{eqnarray}
L_0({\cal Z}) &=& 1
 \nonumber \\
L_1({\cal Z}) &=& \frac{1}{3} \, p_1
 \nonumber \\
L_2({\cal Z}) &=& \frac{1}{45} (-p_1^2 + 7 \, p_2)
 \nonumber \\
L_3({\cal Z}) &=& \frac{1}{945} (2 \, p_1^3 - 13 \, p_1 \, p_2 + 62 \, p_3)
 \nonumber \\
L_4({\cal Z}) &=&
  \frac{1}{14175} (-3 \, p_1^4 + 22\, p_1^2 \, p_2 - 71 \, p_1 \, p_3 - 19 \, p_2^2 + 381 \, p_4)
 \nonumber \\
&\vdots&
\label{Hirzexp}
\end{eqnarray}

The relation (\ref{Hbsigth}) contains additional information, which however has to be extracted case by case for each $n$. For $n=2$ (in which case the base manifold ${\cal Z}$ is 4-dimensional) from 
(\ref{Pontexp}), (\ref{Hbsigth}) and (\ref{Hirzexp}) we obtain
\begin{equation} \label{hirzn2}
\frac{1}{2(2\pi)^2} \int_{\cal Z} \tr(\mathbf{R}^2) = - P_1({\cal Z}) = - 3 \, \tau({\cal Z}) \in 3\,  {\mathbb Z}
\end{equation}
For $n=4$ (8-dimensional ${\cal Z}$) from (\ref{Hbsigth}) and (\ref{Hirzexp}) it follows
\begin{equation}
P_1^2 = 7 \, P_2 \; (\mbox{mod } 45)
\end{equation}
By using this in (\ref{Pontexp}) we obtain
\begin{equation} \label{hirzn4}
\frac{1}{2(2\pi)^4} \int_{\cal Z} \tr(\mathbf{R}^4) =  P_1^2({\cal Z}) - 2 \,P_2({\cal Z})
 = 5 \,P_2({\cal Z}) \; (\mbox{mod } 45) \, \in \, 5 \, {\mathbb Z}
\end{equation}
For $n=6$ we obtain
\begin{equation} \label{hirzn6}
\frac{1}{2(2\pi)^6} \int_{\cal Z} \tr(\mathbf{R}^6) \, \in \, 7\,  {\mathbb Z}
\end{equation}
For $n>6$ the calculations become rapidly more involved, so we stop here.
We see that (\ref{hirzn2}), (\ref{hirzn4}) and (\ref{hirzn6}) are obviously stronger then 
(\ref{inttrirZ}).

\bigskip

Let us now assume that base manifold ${\cal Z}$ can accommodate fermions. Then ${\cal Z}$ must be a spin manifold and we can use Atiyah-Singer index theorem for the Dirac operator to get
\begin{equation} \label{astheor}
\int_{\cal Z} \hat{A}(T{\cal Z}) = \nu_+ - \nu_- \in {\mathbb Z}
\end{equation}
It can be shown that if $n = 2$ (mod 4) the integral above is an \emph{even} integer. $\hat{A}$ is the Dirac genus (or $\hat{A}$-genus) which can also be expressed as a polynomial of the Pontryagin classes $p_j$. Again, there is no general closed form formula, so we list the first few terms of the expansion
\begin{equation} \label{aroofexp}
\hat{A} = 1 - \frac{1}{24} \, p_1 + \frac{1}{5760} (7 \, p_1^2 - 4 \, p_2)
 + \frac{1}{967680} (-31 \, p_1^3 + 44 \, p_1 \, p_2 -16 \, p_3) + \cdots
\end{equation}
The  Atiyah-Singer index theorem (\ref{astheor}) is provides additional information, which however has to be analyzed for each $n$ case by case. For the simplest case $n=2$ using (\ref{astheor}) and (\ref{aroofexp}) we obtain
\begin{equation} \label{asthn2}
\frac{1}{2(2\pi)^2} \int_{\cal Z} \tr(\mathbf{R}^2) = - P_1({\cal Z})
 = 24 \int_{\cal Z} \hat{A}(T{\cal Z}) =  24 \, (\nu_+ - \nu_-) \in 48\,  {\mathbb Z}
\end{equation}
where in addition we used that $n=2$ satisfies $n = 2$ (mod 4), so the left hand side in (\ref{astheor}) is an even integer. In the case $n=4$ from (\ref{astheor}) and (\ref{aroofexp}) follows
\begin{equation}
P_2({\cal Z}) = \frac{7}{4} \, P_1^2({\cal Z}) \; (\mbox{mod } 1440)
\end{equation}
As $P_1^2 ,P_2({\cal Z}) \in {\mathbb Z}$, this relation implies
\begin{equation}
P_1^2({\cal Z}) = 4 \, k \;\;, \qquad P_2({\cal Z}) = 7 \, k \; (\mbox{mod } 1440) \;\;, \qquad k \in {\mathbb Z}
\end{equation}
We can use this to conclude
\begin{equation} \label{asthn4}
\frac{1}{2(2\pi)^4} \int_{\cal Z} \tr(\mathbf{R}^4) =  P_1^2({\cal Z}) - 2 \,P_2({\cal Z})
  \, \in \, 10 \, {\mathbb Z}
\end{equation}
Obviously, (\ref{asthn2}) and (\ref{asthn4}) are stronger conditions then (\ref{hirzn2}) and (\ref{hirzn4}), respectively.

\subsection{CS entropy term}
\label{sapp:gCSent}

In Sec. \ref{ssec:nthorg}, when analyzing consistency of the CS entropy formula in the Euclidean regime, we were faced with (\ref{topintegir}), where the base manifold $\mathcal{Z}$ is now the bifurcation surface 
$\mathcal{B}$ of dimension $2m=2(n-2) \in 4 {\mathbb N}$, $G = U(1)$ and the associated vector bundle is a line bundle (a circle bundle if we use the $G=SO(2)$ formulation). This is a complex bundle for which we can use Chern classes to obtain information about (\ref{topintegir}). The total Chern class is defined by
\begin{equation} \label{tchcldef}
c(\mathbf{F}) \equiv \det \left( 1 + \frac{\mathbf{F}}{2\pi} \right) = \sum_{j=0}^n c_j(\mathbf{F})
\end{equation}
where $c_j(\mathbf{F})$, $j$-th Chern class, is a $2j$-form represents an element of cohomology group $H^{2j}(\mathcal{B})$. By using (\ref{tchcldef}) one can write Chern classes as polynomials of traces of products of $\mathbf{F}$. Explicitly
\begin{eqnarray*}
c_0(\mathbf{F}) &=& 1  \\
c_1(\mathbf{F}) &=& \frac{1}{2\pi} \tr(\mathbf{F}) \\
c_2(\mathbf{F}) &=& \frac{1}{2(2\pi)^2} \left[ (\tr(\mathbf{F}))^2 - \tr(\mathbf{F}^2) \right] \\
&\vdots& \\
c_m(\mathbf{F}) &=& \frac{1}{(2\pi)^m} \det(\mathbf{F})
\end{eqnarray*}
An important property is that Chern numbers, defined as integrals over the whole base manifold 
$\mathcal{B}$ of products of Chern classes with total weight $2m$, are always \emph{integers} if 
$\mathcal{B}$ is an oriented closed manifold
\begin{equation} \label{chnumb}
C_{j_1 \cdots j_r}^{k_1 \cdots k_r}(E) \equiv \int_\mathcal{Z}
 (c_{j_1}(\mathbf{F}))^{k_1} \cdots (c_{j_r}(\mathbf{F}))^{k_r} \, \in \, {\mathbb Z}
 \;\;, \qquad k_1 j_1 + \ldots + k_r j_r = m
\end{equation}
Let us now focus on our problem. We have a line bundle, so all Chern classes except the first vanish
\begin{equation}
c_j(\mathbf{F}) = 0 \;\; \qquad \mbox{for} \quad j > 1
\end{equation}
This also implies that only potentially non-zero Chern number is $C_1^m$.
As the structure group is $U(1)$, in our normalization (which differs by a imaginary unit factor from the usual one in the literature, see in \cite{Nakahara:2003nw,Eguchi:1980jx}) $\mathbf{F}$ is real.
The integral we are interested in is
\begin{equation}
\frac{1}{(2\pi)^m} \int_\mathcal{B} \mathbf{F}^m = C_1^m(E) \, \in \, {\mathbb Z}
\end{equation}
Obviously, if the cohomology group $H^2(\mathcal{B})$ is trivial, then $C_1^m(E) = 0$.

\section{Kruskal-type coordinates}
\label{app:Kruskal}

Here we prove some of the relations and properties used in Section  \ref{app:covphasespace}. The strategy we use is to first make calculations in special ``Kruskal-type" coordinates. Generalization to other coordinate systems typically used in black hole calculations can be done in the same fashion as was done in \cite{BCDPS1}. 

In \cite{Racz:1992bp} it was shown that, in a spacetime with Killing horizon on which the surface gravity is constant, one can construct Kruskal-type coordinates $(U,V,\{x^i\})$, $i=1,\ldots,D-2$, in which metric has the following form
\be \label{kruskal}
ds^2 = G\, dU dV + V H_i\, dx^i dU + g_{ij}\, dx^i dx^j
\ee
where $G$, $H_i$ and $g_{ij}$ are generally smooth functions of $D-1$ variables $U,V$ and $\{x^i\}$. 
The physical horizon is at $U=0$, while $U=V=0$ defines the bifurcation surface $\mathcal{B}$. 
On the bifurcation surface $\mathcal{B}$ we have $G|_{\mathcal{B}} = -2/\kappa$ 
where $\kappa$ is the surface gravity and is constant throughout $\mathcal{B}$. 
We see that 
$\{x^i\}$ are tangential and $U,V$ are normal on $\mathcal{B}$.  The horizon generating Killing vector field 
$\xi$ is given by
\be \label{xiKrus}
\xi = \kappa \left( U \frac{\partial}{\partial U} - V \frac{\partial}{\partial V} \right)
\ee
where the constant $\kappa$ is surface gravity. In Kruskal coordinates, the components of the metric 
(\ref{kruskal}) are regular and well-defined on $\mathcal{B}$, and the components of $\xi$ obviously 
satisfy
\be
\xi^\mu \big|_\mathcal{B} = 0 \quad , \qquad
 \nabla_\nu \xi^\mu \big|_\mathcal{B} = \partial_\nu \xi^\mu \big|_\mathcal{B}  
\ee
and the nonvanishing components of $\partial_\nu \xi^\mu$ on $\mathcal{B}$ are
\be \label{lambdaKrus}
\partial_U \xi^U \big|_\mathcal{B} = - \partial_V \xi^V \big|_\mathcal{B} = \kappa
\ee
From this it follows that
\be \label{dLKrus}
\left( \partial_i \partial_\nu \xi^\mu \right) \big|_\mathcal{B} = 0
\ee

We need also the vielbein in Kruskal coordinates. On $\mathcal{B}$ metric (\ref{kruskal})
is
\begin{equation} \label{KrusB}
ds^2 \big|_\mathcal{B} \equiv g_{\mu\nu}(U=V=0,\{x^i\}) \, dx^\mu dx^\nu
 = - \frac{2}{\kappa} \, dU dV + g_{ij}(\{x^i\}) \, dx^i dx^j
\end{equation}
For practical purposes it is convenient to work in ``light-cone'' basis in which flat Minkowski indices are
$a \in \{u,v,\{ i' \} \}$ in which the non-vanishing components of Minkowski metric $\eta_{ab}$ are
\begin{equation} \label{Minkmet}
\eta_{uv} = - 1 \;\;, \qquad \eta_{i' i'} = 1 
\end{equation}
Using (\ref{KrusB}) and (\ref{Minkmet}) we obtain that nonvanishing components of vielbein $E^a{}_\mu$ on $\mathcal{B}$ are
\begin{equation} \label{Evud}
E^u{}_V \big|_\mathcal{B} = E^v{}_U \big|_\mathcal{B} = \frac{1}{\sqrt{\kappa}} \;\;,
 \qquad E^{i'}{}^j 
\end{equation}
As for the ``inverted'' vielbein $E_a{}^\mu$, nonvanishing components on $\mathcal{B}$ are
\begin{equation} \label{Evdu}
E_u{}^V \big|_\mathcal{B} = E_v{}^U \big|_\mathcal{B} = \sqrt{\kappa} \;\;, \qquad E_{i'}{}^j
\end{equation}
We shall also need $E^a{}^\mu$, for which nonvanishing components on $\mathcal{B}$ are
\begin{equation} \label{Evuu}
E^u{}^U \big|_\mathcal{B} = E^v{}^V \big|_\mathcal{B} =  - \sqrt{\kappa} \;\;, \qquad E^{i'}{}^j
\end{equation}

We now apply this to $L^{ab}$ defined in (\ref{Lab}). Using (\ref{lambdaKrus}) and (\ref{Evud})-(\ref{Evuu}) inside (\ref{Lab}) we obtain that the only nonvanishing components of $L^{ab}$ on $\mathcal{B}$ are 
\begin{equation}
L^{uv} \big|_\mathcal{B} = - L^{vu} \big|_\mathcal{B} = - \kappa = \mbox{const}
\end{equation}
From this obviously follows
\begin{equation} \label{dLabB}
(d L^{ab})_i \big|_\mathcal{B} \equiv (\partial_i L^{ab}) \big|_\mathcal{B} = 0
\end{equation}
A consequence of this result is that in all forms which contain factor of $dL$ when integrated over 
$\mathcal{B}$ give zero. For example, from (\ref{dLabB}) directly follows that
\begin{equation} \label{iBSig0}
\int_\mathcal{B} \Sig_{\xi,L} = 0
\end{equation}
where $\Sig_{\xi,L}$ is defined in (\ref{sigmaL}).

An important consequence following from (\ref{Evud})-(\ref{Evuu}) is that in Kruskal coordinates
\begin{equation} \label{dEB}
(d E)_i \big|_\mathcal{B} \equiv (\partial_i E) \big|_\mathcal{B} = 0
\end{equation}
Used in (\ref{Gammaalpha}) this implies
\begin{equation} \label{GamalphaB}
\Gamma_i \big|_\mathcal{B} = E^{-1} \hat{\alpha}_i \, E \big|_\mathcal{B}
\end{equation}
This in turn can be used to show that the formula for the entropy (\ref{SCSpta}) is (locally) the same as the one we derived in \cite{BCDPS1}, Eq.\ (4.11).

\vskip 2cm



\begin{thebibliography}{99}

\bibitem{Tachi}
  Y.~Tachikawa,
  {\it Black hole entropy in the presence of Chern-Simons terms},
  Class.\ Quant.\ Grav.\  {\bf 24} (2007) 737 \
  [\texttt{arXiv:hep-th/0611141}].

\bibitem{BCDPS1}
  L.~Bonora, M.~Cvitan, P.~Dominis Prester, S.~Pallua and I.~Smoli\'{c},
  {\it Gravitational Chern-Simons Lagrangians and black hole entropy},
  JHEP {\bf 1107} (2011) 085 \
  [\texttt{arXiv:1104.2523 [hep-th]}].

\bibitem{Solodukhin:2005ns}
  S.~N.~Solodukhin,
  {\it Holographic description of gravitational anomalies},
  JHEP {\bf 0607} (2006) 003 \
  [\texttt{arXiv:hep-th/0512216}].

\bibitem{BCDPS2}
  L.~Bonora, M.~Cvitan, P.~D.~Prester, S.~Pallua and I.~Smoli\'{c},
  {\it Gravitational Chern-Simons Lagrangian terms and spherically symmetric spacetimes},
  Class.\ Quant.\ Grav.\  {\bf 28} (2011) 195009 \
  [\texttt{arXiv:1105.4792 [hep-th]}].

\bibitem{Wald1}
  R.M.~Wald,
  {\it Black hole entropy is the Noether charge,}
  Phys.~Rev.~D {\bf 48} (1993) R3427 \ 
  [\texttt{arXiv:gr-qc/9307038v1}]

\bibitem{Iyer}
  V.~Iyer and R.~M.~Wald,
  ``Some properties of Noether charge and a proposal for dynamical black hole entropy,''
  Phys.\ Rev.\ D {\bf 50} (1994) 846 \
  [\texttt{arXiv:gr-qc/9403028}].

\bibitem{JKM}
T.~Jacobson, G.~Kang and R.C.~Myers,
  {\it On black hole entropy,}
  Phys.~Rev.~D {\bf 49} (1994) 6587 \
  [\texttt{arXiv:gr-qc/9312023v2}]

\bibitem{Wald2} 
R.M.~Wald, {\it On identically closed forms locally constructed from a field},
  Jour.~Math.~Phys. 31 (1990) 2378.

\bibitem{Wald3}
  R.~M.~Wald, A.~Zoupas,
  {\it A General definition of 'conserved quantities' in general relativity and other theories of gravity,}
  Phys.\ Rev.\  {\bf D61 } (2000)  084027 \
  [\texttt{arXiv:gr-qc/9911095}].

\bibitem{Racz1}
  I.~Racz and R.~M.~Wald,
  {\it Global extensions of space-times describing asymptotic final states of black holes},
  Class.\ Quant.\ Grav.\  {\bf 13} (1996) 539 \
  [\texttt{arXiv:gr-qc/9507055}].

\bibitem{Racz2}
  I.~Racz and R.~M.~Wald,
  {\it Extension of space-times with Killing horizon},
  Class.\ Quant.\ Grav.\  {\bf 9} (1992) 2643.

\bibitem{Witten}
  E.~Witten,
  ``Three-Dimensional Gravity Revisited,''
  [\texttt{arXiv:0706.3359 [hep-th]}].

\bibitem{W}
  E.~Witten,
  {\it Quantization of Chern-Simons gauge theory with complex gauge group},
  CMP {\bf 137} (1990) 29.

\bibitem{DW}
  R.~Dijkgraaf and E.~Witten,
  {\it Topological gauge theories and group cohomology},
  CMP {\bf 129} (1990) 393.
 
\bibitem{DJT1}
  S.~Deser, R.~Jackiw and S.~Templeton,
  {\it Three-dimensional massive gauge theories,}
  Phys.~Rev.~Lett. {\bf 48} (1982) 975.

\bibitem{DJT2}
  S.~Deser, R.~Jackiw and S.~Templeton,
  {\it Topologically massive gauge theories,}
  Ann.~Phys., NY {\bf 140} (1982) 372.
 
\bibitem{Witten2}
  E.~Witten,
  {\it (2+1) Dimensional Gravity as an Exactly Soluble System}
  Nucl.Phys. {\bf B311} (1988) 46.

\bibitem{Banados}
  M.~Banados, C.~Teitelboim and J.~Zanelli,
  ``The Black hole in three-dimensional space-time,''
  Phys.\ Rev.\ Lett.\  {\bf 69} (1992) 1849 \
  [\texttt{arXiv:hep-th/9204099}].

\bibitem{DT}
  S.~Deser and B.~Tekin,
  {\it Energy in topologically massive gravity,}
  Class.~Quantum Grav. {\bf 20} (2003) L259 \
  [\texttt{arXiv:gr-qc/0307073v1}]

\bibitem{WittenM}
  A.~Maloney and E.~Witten,
 {\it Quantum Gravity Partition Functions in Three Dimensions,}
  JHEP {\bf 1002} (2010) 029 \
  [\texttt{arXiv:0712.0155 [hep-th]}].

\bibitem{SolodukhinCS}
  S.~N.~Solodukhin,
  {\it Holography with gravitational Chern-Simons,}
  Phys.\ Rev.\  {\bf D74 } (2006)  024015 \
  [\texttt{arXiv:hep-th/0509148}].

\bibitem{Perez}
  R.~F.~Perez,
  {\it Conserved current for the Cotton tensor, black hole entropy and equivariant Pontryagin forms,}
  Class.\ Quant.\ Grav.\  {\bf 27 } (2010)  135015 \
  [\texttt{arXiv:arXiv:1004.3161 [gr-qc]}].

\bibitem{Kraus}
  P.~Kraus, F.~Larsen,
  {\it Holographic gravitational anomalies,}
  JHEP {\bf 0601 } (2006)  022 \
  [\texttt{arXiv:hep-th/0508218}].

\bibitem{Park}
  M.~-I.~Park,
  {\it BTZ black hole with gravitational Chern-Simons: Thermodynamics and statistical entropy,}
  Phys.\ Rev.\  {\bf D77 } (2008)  026011 \
  [\texttt{arXiv:hep-th/0608165}].

\bibitem{Miskovic}
  O.~Mi\v{s}kovi\'{c}, R.~Olea,
  {\it Background-independent charges in Topologically Massive Gravity,}
  JHEP {\bf 0912 } (2009)  046 \
  [\texttt{arXiv:0909.2275 [hep-th]}].

\bibitem{CS}
  S.~Chern and J.~Simons,
  {\it Characteristic forms and geometric invariants},
  Ann.\ Math. {\bf 99} (1974) 48.

\bibitem{KN}
  S.~Kobayashi and K.~Nomizu, 
  {\it Foundation of differential geometry} , vol.I, II,  
  John Wiley \& Sons, New York 1963.

\bibitem{Carter:2000wv}
  B.~Carter,
  {\it Essentials of classical brane dynamics},
  Int.\ J.\ Theor.\ Phys.\  {\bf 40} (2001) 2099 \
  [\texttt{arXiv:gr-qc/0012036}].

\bibitem{BCDPSp}
  L.~Bonora, M.~Cvitan, P.~Dominis Prester, S.~Pallua and I.~Smoli\'{c},
  ``Symmetries and gravitational Chern-Simons Lagrangian terms'',
  in preparation.

\bibitem{MS}
  J.W.~Milnor and J.D.~Stasheff, 
  {\it Characteristic classes}, 
  Annals of Mathematics Studies, No. 76. Princeton University Press, Princeton, N. J.;
  University of Tokyo Press, Tokyo, 1974. 

\bibitem{Stong} 
  R.E.~Stong, 
  {\it Notes on cobordism theory}, 
  Princeton Univ. Press, (1968).

\bibitem{WittenC}
  E.~Witten,
  {\it 	On flux quantization in M theory and the effective action},
  J.\ Geom.\ Phys. {\bf 22} (1997) 1.

\bibitem{Lu:2010sj}
  H.~Lu and Y.~Pang,
  {\it Seven-Dimensional Gravity with Topological Terms},
  Phys.\ Rev.\ D {\bf 81} (2010) 085016 \
  [\texttt{arXiv:1001.0042 [hep-th]}].

\bibitem{Galloway}
  G.~J.~Galloway,
  {\it Constraints on the topology of higher dimensional black holes},
  \texttt{arXiv:1111.5356 [gr-qc]}.

\bibitem{HoIsh}
  S.~Hollands and A.~Ishibashi,
  {\it Black hole uniqueness theorems in higher dimensional spacetimes},
  Class.\ Quant.\ Grav.\  {\bf 29} (2012) 163001 \
  [arXiv:1206.1164 [gr-qc]].

\bibitem{KlKuRa} 
  B.~Kleihaus, J.~Kunz and E.~Radu,
  {\it Black rings in more then five dimensions},
  \texttt{arXiv:1205.5437 [hep-th]}.

\bibitem{KlKuRaRo}
  B.~Kleihaus, J.~Kunz, E.~Radu and M.~J.~Rodriguez,
  JHEP {\bf 1102 } (2011)  058 \
  \texttt{arXiv:1010.2898 [gr-qc]}.

\bibitem{Schwartz}
  F.~Schwartz,
  {\it Existence of outermost apparent horizons with product of spheres topology},
  Commun.\ Anal.\ Geom.\  {\bf 16} (2008) 799 \
  \texttt{arXiv:0704.2403 [gr-qc]}.

\bibitem{BY} 
  J.~D.~Brown and J.~W.~York, Jr.,
  {\it The Path integral formulation of gravitational thermodynamics,}
  In *Teitelboim, C. (ed.) et al.: The black hole* 1-24
  [\texttt{arXiv:gr-qc/9405024}].

\bibitem{BZ}
  W.A.~Bardeen and B.~Zumino, 
  {\it Consistent and covariant anomalies in gauge and gravitational theories}, 
  Nucl.Phys. {\bf B244} (1984) 421.

\bibitem{AlvarezGaume:1984dr}
  L.~Alvarez-Gaume and P.~H.~Ginsparg,
  {\it The Structure of Gauge and Gravitational Anomalies},
  Annals Phys.\  {\bf 161} (1985) 423
   [Erratum-ibid.\  {\bf 171} (1986) 233].

\bibitem{Eguchi:1980jx}
  T.~Eguchi, P.~B.~Gilkey and A.~J.~Hanson,
  {\it Gravitation, Gauge Theories and Differential Geometry},
  Phys.\ Rept.\  {\bf 66} (1980) 213.

\bibitem{Nakahara:2003nw}
  M.~Nakahara,
  {\it Geometry, topology and physics},
  Boca Raton, USA: Taylor \& Francis (2003)

\bibitem{Racz:1992bp}
  I.~Racz and R.~M.~Wald,
  {\it Extension of space-times with Killing horizon},
  Class.\ Quant.\ Grav.\  {\bf 9} (1992) 2643.
  

\end{thebibliography}
\end{document}